\documentclass[aps,prx,twocolumn,showpacs]{revtex4-1}
\usepackage{bm}
\usepackage{mathrsfs}
\usepackage{amsmath}
\usepackage{amssymb}
\usepackage{graphicx}
\usepackage{amsfonts}
\usepackage{amsthm}
\usepackage{color}
\usepackage{dcolumn}
\usepackage{txfonts}
\usepackage[caption=false]{subfig}
\usepackage[T1]{fontenc}
\usepackage{multirow}

\DeclareMathOperator{\cov}{cov}
\DeclareMathOperator\erf{erf}

\begin{document}

\title{Fast and high-fidelity dispersive readout of a spin qubit with squeezing and resonator nonlinearity}
\author{Chon-Fai Kam}
\email{Email: dubussygauss@gmail.com}
\affiliation{Department of Physics, University at Buffalo, SUNY, Buffalo, New York 14260, USA}
\author{Xuedong Hu}
\email{Email: xhu@buffalo.edu}
\affiliation{Department of Physics, University at Buffalo, SUNY, Buffalo, New York 14260, USA}

\begin{abstract}
Fast and high-fidelity qubit measurement is crucial for achieving quantum error correction, a fundamental element in the development of universal quantum computing. For electron spin qubits, fast readout stands out as a major obstacle in the pursuit of error correction. In this work, we explore the dispersive measurement of an individual spin in a semiconductor double quantum dot coupled to a nonlinear microwave resonator. By utilizing displaced squeezed vacuum states, we achieve rapid and high-fidelity readout for semiconductor spin qubits. Our findings reveal that introducing modest squeezing and mild nonlinearity can significantly improve both the signal-to-noise ratio (SNR) and the fidelity of qubit-state readout. By properly marching the phases of squeezing, the nonlinear strength, and the local oscillator, the optimal readout time can be reduced to the sub-microsecond range. With current technology parameters ($\kappa\approx 2\chi_s$, $\chi_s\approx 2\pi\times 0.15 \:\mbox{MHz}$), utilizing a displaced squeezed vacuum state with $30$ photons and a modest squeezing parameter $r\approx 0.6$, along with a nonlinear microwave resonator charactered by a strength of $\lambda\approx -1.2 \chi_s$, a readout fidelity of $98\%$ can be attained within a readout time of around $0.6\:\mu\mbox{s}$. Intriguing, by using a positive nonlinear strength of $\lambda\approx 1.2\chi_s$, it is possible to achieve an SNR of approximately $6$ and a readout fidelity of $99.99\%$ at a slightly later time, around $0.9\:\mu\mbox{s}$, while maintaining all other parameters at the same settings.
\end{abstract}

\maketitle

\section{Introduction}
The pursuit of constructing a large-scale universal quantum computer, the prime goal in the field of quantum information science, presents an immensely challenging task that necessitates advancements in both quantum algorithms and hardware development. Currently, there are multiple approaches to quantum computing, including superconducting circuits \cite{kjaergaard2020superconducting, chen2021exponential}, trapped ions \cite{bruzewicz2019trapped, wright2019benchmarking}, semiconductor spin qubits \cite{burkard2021semiconductor, noiri2022fast, xue2022quantum, mkadzik2022precision, takeda2022quantum}, and photonic systems \cite{takeda2019toward, bourassa2021blueprint, madsen2022quantum}. Up to now, the largest superconducting quantum computers have up to a few hundreds of qubits \cite{riel2022quantum}, while the largest trapped ion quantum computers has scaled up to a few tens of qubits under coherent control \cite{egan2021fault}. Among these options, spin qubits in silicon-based semiconductor quantum dots, while lagging somewhat behind other approaches in early qubit demonstrations, offers tantalizing long-term potential to combine the high controllability of trapped ion systems as well as the scalability of superconducting systems. Spin qubits in semiconductor quantum dots hold great promise as a quantum computing platform owing to their scalability infrastructure rooted in a compact qubit footprint of roughly $0.01\:\mu\mbox{m}^2$, along with their compatibility with the widely-established complementary metal-oxide-semiconductor (CMOS) integration techniques developed in the last half-century \cite{gonzalez2021scaling}. 

Nonetheless, in the pursuit of scalable quantum computing, a pivotal obstacle regardless of the qubit's physical nature, lies in successfully incorporating active quantum error correction for logical qubits while preserving the encoded information \cite{cramer2016repeated}. This entails the need for stabilizer measurements of multiple qubits \cite{terhal2015quantum}, along with the implementation of real-time feedback to the encoded qubits \cite{saraiva2022dawn}. The effectiveness of real-time feedback control for active quantum error correction depends on the capability to attain high-fidelity readout within timescales considerably shorter than the qubits' decoherence times. In silicon quantum dots, spin coherence times typically range from microseconds to milliseconds, depending on the specific design and control methods \cite{vandersypen2019quantum}, making it imperative to achieve readout fidelity surpassing the 99\% threshold required by the most prevalent surface error correction codes \cite{raussendorf2007fault, fowler2012surface} in times of a few microseconds. Although gate-based dispersive spin readout in silicon has shown an average readout fidelity ranging from 73.3\% to 98\% when utilizing off-chip resonators, it necessitates integration times on the order of milliseconds (0.3 ms - 2.6 ms) to accomplish single-shot readout \cite{pakkiam2018single, west2019gate, urdampilleta2019gate}. 

Spin measurement approaches commonly used, such as those relying on spin-dependent tunneling \cite{elzerman2004single} or spin blockade \cite{petta2005coherent}, along with a DC charge sensor, are too slow for active quantum error correction. Employing on-chip resonators with a higher-frequency probe pulse, a gate-based dispersive spin readout in silicon has yielded an average readout fidelity of 98\%, achieving single-shot readout in a measurement time of 6 $\mu$s \cite{zheng2019rapid}. Besides, by employing a single-electron transistor (SET) as a charge sensor has achieved a single-shot readout for electron spins in silicon with a fidelity of 97\% within 1.5 $\mu$s \cite{keith2019single}, as well as a fidelity of 99.9\% within 6 $\mu$s \cite{curry2019single}. Moreover, by utilizing a radio-frequency single-electron transistor (rf-SET), researchers have achieved a signal-to-noise ratio (SNR) of 6 for a single-shot singlet-triplet readout within 0.8 $\mu$s \cite{noiri2020radio}, and 99\% fidelity within 1.6 $\mu$s \cite{connors2020rapid}. Nonetheless, integrating charge sensors would considerably increase the complexity of the device, while the larger dimensions of rf-SETs limit their positioning within densely connected qubit frameworks \cite{de2023silicon}. Consequently, in the context of quantum dot spin qubits, readout processes remain notably slower than gate operations when on-chip charge sensors are not employed. In this regard, achieving a fast readout with high fidelity and a compact footprint for quantum dot spin qubits remains a formidable challenge, acting as a bottleneck in the advancement of current quantum information technology based on spin qubits.

In order to address the readout bottleneck associated with spin qubits, it becomes imperative to enhance the readout sensitivity into the sub-microsecond time regimes. One approach can be achieved in dispersive readout by employing squeezed microwaves as an alternative to the conventional coherent states. Squeezed states are nonclassical states with modified quantum noise profiles \cite{walls1983squeezed, zhang1990coherent, kam2023coherent}. By introducing quantum correlations among photons, quantum fluctuations in one field quadrature can be substantially reduced below the coherent-state level, at the expense of amplifying fluctuations in the conjugate one \cite{andersen201630}. Squeezed states have undergone thorough examination across multiple research domains in the past decades. Notably, the utilization of squeezed states of light has played a pivotal role in recent laser gravitational wave experiments conducted with second-generation detectors \cite{tse2019quantum, acernese2019increasing}, which leads to a nearly tenfold enhancement in gravitational wave detection sensitivity \cite{dwyer2022squeezing}. Furthermore, they have found applications in diverse areas such as continuous-variable quantum key distribution \cite{gehring2015implementation}, quantum sensing \cite{lawrie2019quantum}, high-precision cavity spectroscopy \cite{junker2021high}, as well as Bose-Einstein condensates \cite{lyu2020geometrizing} and biosensing \cite{li2020squeezed}.

In addition to their fundamental significance as nonclassical states, squeezed states of light have found practical application in improving the SNR of superconducting qubit readouts in the domain of quantum computing \cite{barzanjeh2014dispersive}. When employed optimally, squeezed states can push towards the Heisenberg scaling limit, in which the SNR scales not linearly with the number of photons but rather as the square of that number. Nonetheless, realizing exponential growth in SNR and attaining the Heisenberg scaling limit necessitates unconventional longitudinal interactions between the qubit and oscillator \cite{didier2015fast}, or a pair of readout modes with unconventional dispersive couplings that are differ in sign \cite{didier2015heisenberg, govia2017enhanced}, as opposed to the standard readout techniques used in dispersive qubit measurements. Additionally, experiments demonstrated that using squeezed microwaves yielded a 24\% increase in the final SNR for superconducting qubit measurements \cite{eddins2018stroboscopic}. Furthermore, the use of a two-mode squeezed vacuum state resulted in a 31\% improvement in the SNR for superconducting qubit readout with 99\% fidelity \cite{liu2022noise}. 

Intriguingly, unlike the dispersive readout scheme which entails the injection of externally prepared squeezed states, there is an alternative approach that utilizes intra-resonator squeezing to enhance qubit readout fidelity. In the realm of superconducting qubit measurements, the inclusion of intra-resonator squeezing has demonstrated a nearly five-fold reduction in the measurement time needed to reach 99.99\% fidelity in the limit of large coherent drive \cite{govia2017enhanced}. In this regard, the utilization of either external or internal squeezing can enhance readout fidelity and reduce the measurement time required for a specific fidelity level. Consequently, one can naturally inquire: Is there still potential for further improving readout fidelity at shorter measurement times when both external and internal squeezing are employed? 

In a previous work, we have demonstrated sub-microsecond high-fidelity dispersive readout of a spin qubit with squeezed photons \cite{kam2023sub}. In this study, we will delve into the impact of both external and internal squeezing on the dispersive readout of a spin qubit coupled to a microwave resonator. Our finding shows that, when specific phase matching conditions are met, utilizing a displaced squeezed vacuum state with moderate squeezing levels along with an intra-resonator nonlinearity (internal squeezing) can lead to substantial enhancements in the SNR, enabling fast and high-fidelity readout of the spin qubit through standard dispersive spin-readout techniques. 

\section{The Model}
To begin with, we consider the dispersive readout of a spin qubit assisted by a single mode microwave resonator. In the following, we will analyze the dynamics of the resonator in a rotating frame of reference with a probe frequency $\omega_{\mathrm{in}}$. In the rotating reference frame, the dispersive Hamiltonian has the form \cite{d2019optimal, govia2017enhanced}
\begin{equation}\label{Dispersive}
    H_s = \frac{1}{2}(\delta_s-\chi_s)\sigma_z+\left(\delta_c-\chi_s\sigma_z\right)a^\dagger a +\frac{i}{2}(\lambda a^{\dagger 2}-\lambda^* a^2),
\end{equation}
where $\delta_c\equiv \omega^\prime_r-\omega_{\mathrm{in}}$ and $\delta_s\equiv E_s^\prime -\omega_{\mathrm{in}}$ are the detunings of the probe from the resonator and spin qubits, respectively, $\omega_r^\prime$ and $E_s^\prime$ are the resonator and spin-qubit frequencies, respectively, and $\chi_s$ is the dispersive coupling strength. In Eq.\:\eqref{Dispersive}, we introduce a second-order nonlinearity in the resonator with $\lambda\equiv \lambda_x+i\lambda_y\equiv |\lambda|e^{i\theta_\lambda}$ denoting the nonlinear strength, resulting in the generation of intra-resonator squeezing. Given that photons can leak in and out of the resonator through the input port with a leakage rate $\kappa$, it becomes necessary to incorporate an interaction term of the input radiation and the resonator \cite{d2019optimal}
\begin{equation}\label{interaction}
V_{\mathrm{in}}=i\sqrt{\kappa}(b_{\mathrm{in}}^\dagger a-b_{\mathrm{in}}a^\dagger),
\end{equation}
where $b_{\mathrm{in}}$ and $b_{\mathrm{in}}^\dagger$ are the annihilation and creation operators of the input radiation field which satisfy the commutation relation $[b_{\mathrm{in}}(t),b^\dagger_{\mathrm{in}}(t^\prime)]=\delta(t-t^\prime)$. Notice that in principle, one can have more than one input radiations via multiple ports, but for simplicity, we herein consider only the case of a single input port. From Eqs.\:\eqref{Dispersive} - \eqref{interaction}, one may derive the effective Hamiltonian $H_{\mathrm{eff}}\equiv H_s+V_{\mathrm{in}}$, and thus the Langevin equation of motion for the resonator field
\begin{equation}\label{Master}
    \dot{a} = -\left[i\left(\delta_c-\chi_s\sigma_z\right)+\frac{\kappa}{2}\right]a +\lambda a^\dagger - \sqrt{\kappa}b_{\mathrm{in}},
  \end{equation}
where the first term describes the dispersive motion of the resonator field in the present of a resonator damping, the second term is a coupling between annihilation and creation field operators due to the intra-resonator second-order nonlinearity, and the last term describes driving of the resonator through its input port. Notice that the damping term in Eq.\:\eqref{Master} is Markovian, i.e., the damping terms depend only on the system operators at an instant time. It comprises the core assumption of the quantum input-output theory of Gardiner and Collet \cite{gardiner1985input, gardiner2004quantum}. 

Under such a Markovian assumption, the output radiation field can be determined by the input-output relation \cite{gardiner1985input}
\begin{equation}\label{io}
b_{\mathrm{out}}=b_{\mathrm{in}}+\sqrt{\kappa}a,
\end{equation}
where $b_{\mathrm{out}}$ and $b_{\mathrm{out}}^\dagger$ are the annihilation and creation operators of the output radiation field which is about to propagate away. One has to bear in mind that the input-output relation Eq.\:\eqref{io} is general. It is valid even for nonlinear systems. In order to evaluate the qubit readout performance, one may perform a quantum non-demolition measurement of the spin qubit, in which $\sigma_z(t)$ is approximately a constant of motion, i.e., $\sigma_z(t)\approx \sigma_z(0)$. As long as the spin qubit is coupled to a relatively weak field, such that the qubit relaxation can be neglected, one may employ this assumption to analyze the readout contrast. In order words, in this scenario, one only needs to consider the qubit in a definite state, and consequently one can write $\sigma_z$ as a real number $\sigma=\pm 1$ from here on. Without loss of generality, one may assume $\delta_c = 0$ onwards.

In RF-reflectometry, the output field $b_{\mathrm{out}}$ is sent through a phase-preserving amplifier \cite{clerk2010introduction}, which equally detects both quadratures. Subsequently, the output field is measured using a homodyne detector \cite{yuen1983noise} by mixing with a local oscillator with phase $\varphi$. The resulting photocurrent $I\propto \frac{1}{\sqrt{2}}(b_{\mathrm{out}}e^{-i\varphi}+b_{\mathrm{out}}^\dagger e^{i\varphi})$, where $\varphi=0$ or $\pi/2$ representing $Q_{\mathrm{out}}$ or $P_{\mathrm{out}}$ respectively. We thus revise the Langevin equations for the resonator field in terms of the quadratures as
\begin{subequations}
\begin{align}\label{Langevin1}
\dot{X}(t)=
M X(t)
&-\sqrt{\kappa}X_{\mathrm{in}},X\equiv(Q,P)^\top,  X_{\mathrm{in}}\equiv(Q_{\mathrm{in}},P_{\mathrm{in}})^\top,\\
M&\equiv\begin{pmatrix}
-\frac{\kappa}{2}+\lambda_x & \mp\chi_s+\lambda_y \\
\pm\chi_s+\lambda_y & -\frac{\kappa}{2}-\lambda_x
\end{pmatrix}\nonumber\\
&\equiv-\frac{\kappa}{2}+\lambda_y\tau_x\mp i \chi_s \tau_y+\lambda_x\tau_z,\label{Langevin2}
\end{align}
\end{subequations}
where $Q\equiv(a^\dagger +a)/\sqrt{2}$ and $P\equiv i(a^\dagger -a)/\sqrt{2}$ are the amplitude and phase quadratures of the resonator field. Similarly, $Q_{\mathrm{in}}\equiv (b^\dagger_{\mathrm{in}} +b_{\mathrm{in}})/\sqrt{2}$ and $P_{\mathrm{in}}\equiv i(b_{\mathrm{in}}^\dagger -b_{\mathrm{in}})/\sqrt{2}$ are the quadratures of the input radiation field. As the Langevin equations of motion, Eqs.\:\eqref{Langevin1} - \eqref{Langevin2}, are linear in the quadrature operators $Q$ and $P$, they can be solved exactly. Assuming the input radiation field is time-independent, i.e., a continuous-wave (CW) radiation, the Langevin equations for the quadratures of the resonator field are solved by
\begin{gather}
X(t) = e^{Mt}X(0)-\sqrt{\kappa}\int_0^tds e^{Ms} X_{\mathrm{in}},
\end{gather}
where
\begin{subequations}
\begin{gather}
    e^{Mt}=f(t) +g(t)\boldsymbol{n}\cdot\boldsymbol{\tau},\\
    f(t)\equiv e^{-\frac{\kappa}{2}t}\cosh\Delta t,g(t)\equiv e^{-\frac{\kappa}{2}t}\sinh\Delta t,\\
    \Delta^2\equiv \lambda_x^2+\lambda_y^2-\chi_s^2,\boldsymbol{n}\equiv \frac{1}{\Delta}(\lambda_y,\mp i\chi_s,\lambda_x).
\end{gather}
\end{subequations}
Here, $\boldsymbol{\tau}$ is the vector Pauli matrix acting on the two quadratures $Q(t)$ and $P(t)$.

\begin{figure}[tbp]
	\subfloat[$\kappa=\chi_s$\label{sfig:1a}]{%
	\includegraphics[width=0.5\columnwidth]{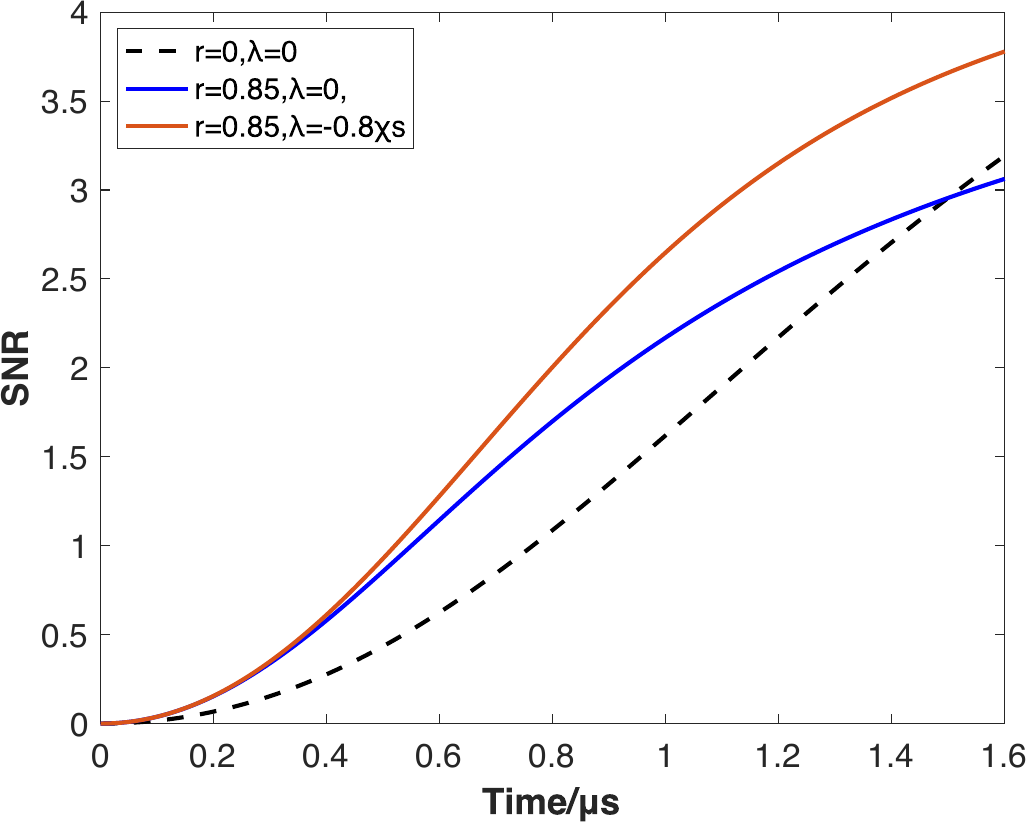}%
	}\hfill
	 \subfloat[$\kappa=\chi_s$\label{sfig:1b}]{%
	\includegraphics[width=0.5\columnwidth]{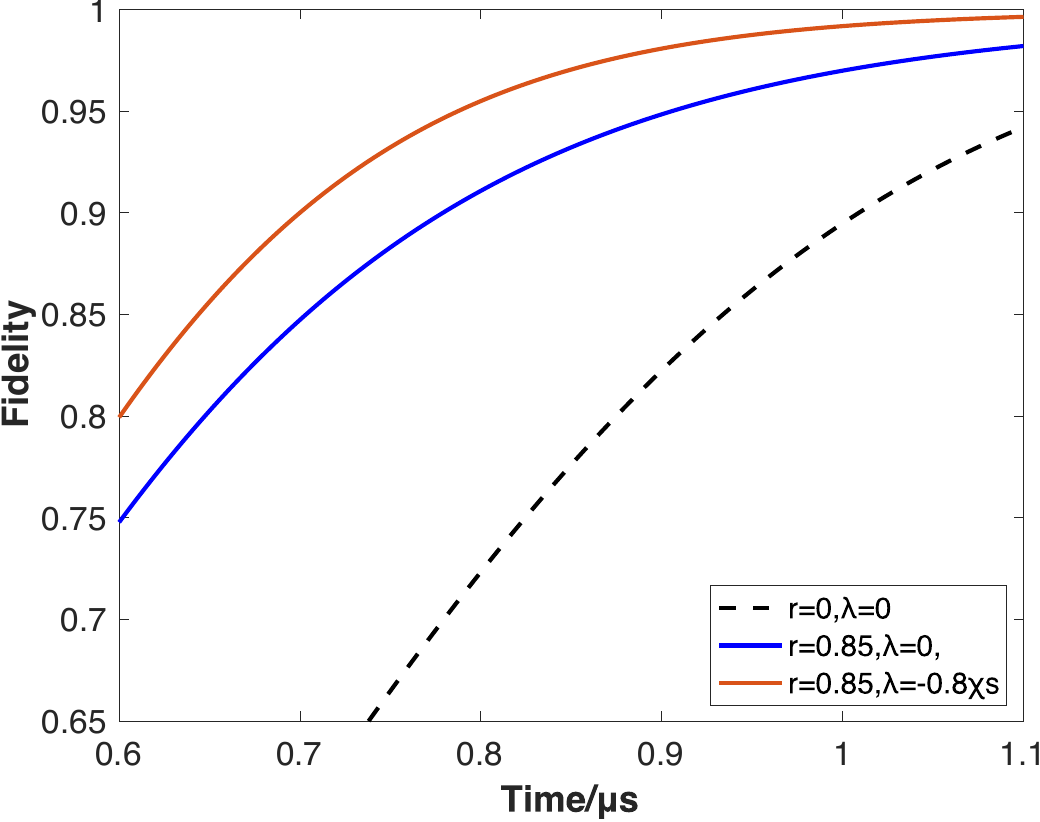}%
	 }\hfill
	\subfloat[$\kappa=2\chi_s$\label{sfig:1c}]{%
 	 \includegraphics[width=0.5\columnwidth]{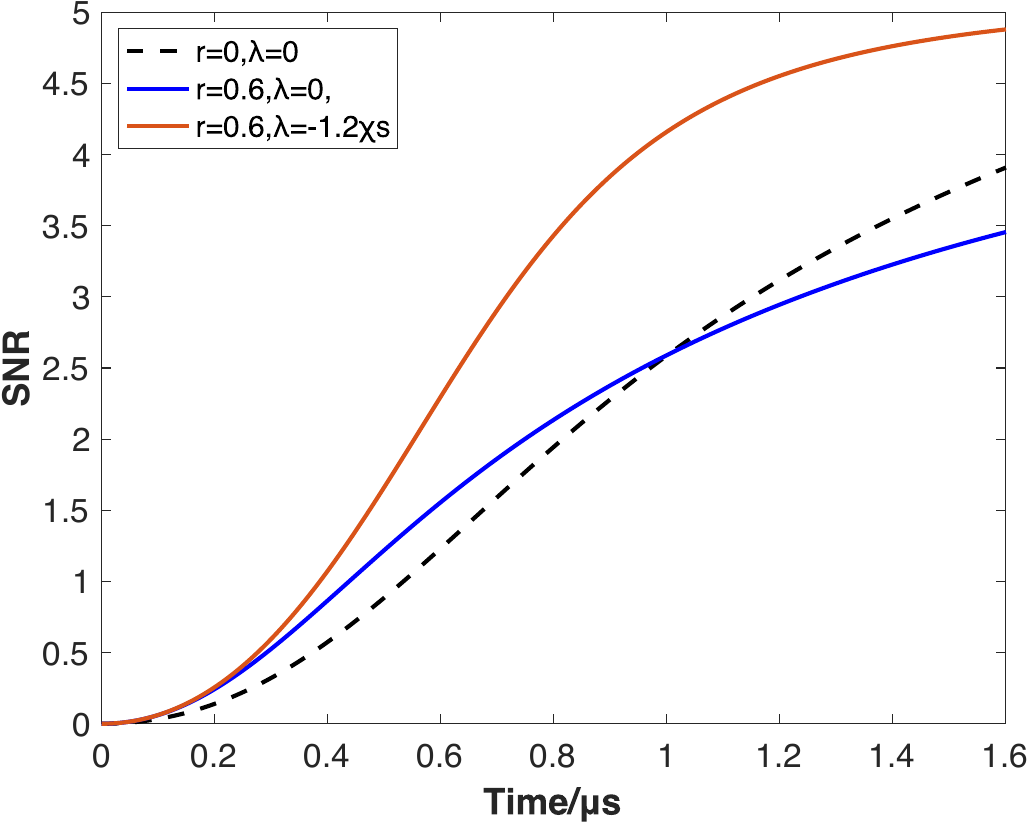}%
	}\hfill
	\subfloat[$\kappa=2\chi_s$\label{sfig:1d}]{%
 	 \includegraphics[width=0.5\columnwidth]{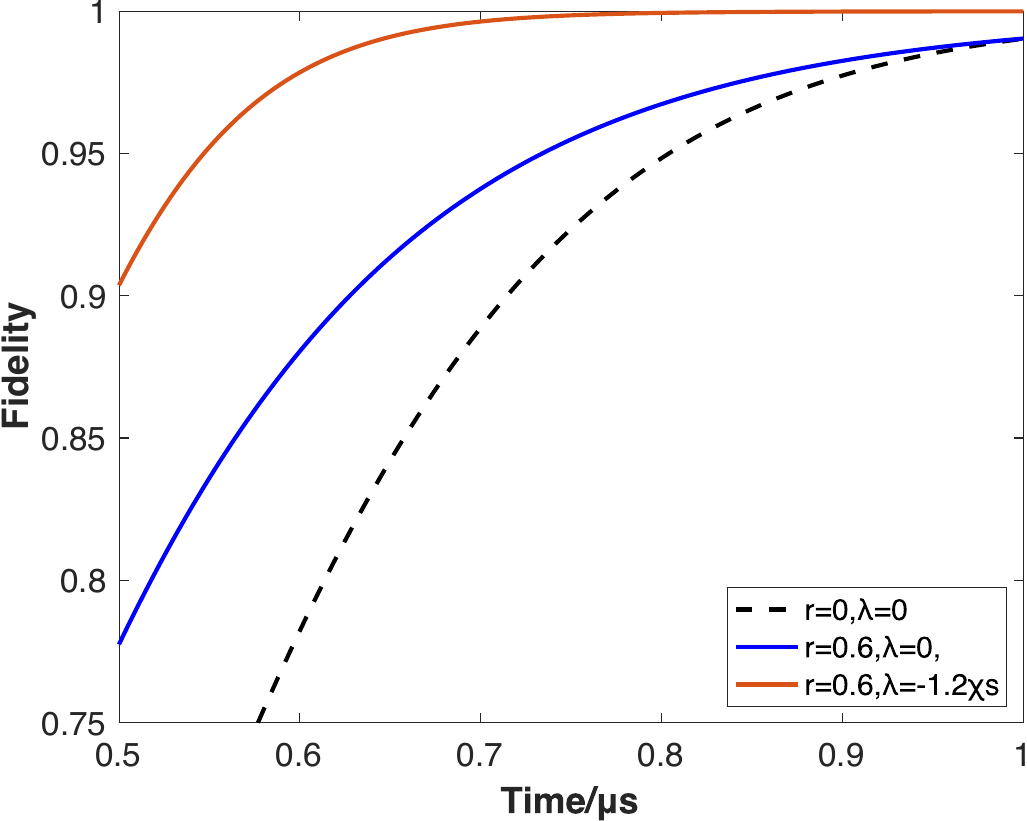}%
	}
\caption{Schematics of the signal-to-noise ratio ($\mbox{SNR}$) and the single-shot readout fidelity with respect to different leakage rates $\kappa$, squeezing parameters $r$, and nonlinear strengths $\lambda$. Here, the dispersive coupling strength, the amplitude of the input displaced squeezed vacuum state, and the spin $T_1$ relaxation time are: $\chi_s=2\pi\times 0.15\:\mbox{MHz}$, $|\alpha|=\sqrt{30}$, and $T_1=10\:\mbox{ms}$. The black dashed lines represent the coherent state inputs.}
\label{fig:Fig1}
\end{figure}

\section{Signal-to-noise ratio \& Fidelity}
\subsection{Signal-to-noise ratio}
To unveil information encoded in qubit states, a standard approach involves performing a homodyne measurement on the resonator field, which yields a homodyne current that is proportional to the expectation of the output quadrature. Particularly, one can fix the output quadrature as $(b_{\mathrm{out}}e^{-i\varphi}+b_{\mathrm{out}}^\dagger e^{i\varphi})/\sqrt{2}=\cos\varphi Q_{\mathrm{out}}+\sin\varphi P_{\mathrm{out}}$ by choosing the local oscillator phase as $\varphi$. To evaluate the distinguishability between the two qubit states, it is valuable to introduce the signal-to-noise ratio (SNR). The SNR serves as a convenient measure for quantifying the degree to which the qubit states can be distinguished through measurement outcomes, which is defined as the ratio of the contrast between the output signals to the sum of the associated standard deviations
\begin{equation}\label{SNRDef}
\mbox{SNR}(t)\equiv\frac{|\langle \mathscr{M}^{+1}(t)\rangle-\langle \mathscr{M}^{-1}(t)\rangle|}{\Delta \mathscr{M}^{+1}(t)+\Delta \mathscr{M}^{-1}(t)}.
\end{equation}
Here, $\langle \mathscr{M}^{\pm}(t)\rangle$ and $\Delta \mathscr{M}^{\pm 1}(t)$ are the expectations and standard derivations of the time-integrated output quadratures respectively
\begin{equation}
\mathscr{M}^{\pm 1}(t)\equiv \cos\varphi\int_0^t Q^{\pm 1}_{\mathrm{out}}(s)ds+\sin\varphi \int_0^t P^{\pm 1}_{\mathrm{out}}(s)ds,
\end{equation}
where the superscript $\pm 1$ is used to indicate the output quadratures corresponding to the two qubit states respectively. To be specific, the time-integrated output quadratures can be explicitly expressed as (see Appx.\:\ref{A})
\begin{align}\label{Signal}
   &\mathscr{M}^\pm(t)\equiv A(t)Q_{\mathrm{in},\varphi}+ \frac{|\lambda|}{\Delta}B(t)Q_{\mathrm{in},\pi-\varphi^\prime}\pm\frac{\chi_s}{\Delta}B(t)P_{\mathrm{in},\varphi}\nonumber\\
   +&\sqrt{\kappa}[F(t)Q_{\varphi}(0)-\frac{|\lambda|}{\Delta}G(t)Q_{\pi-\varphi^\prime}(0)\mp\frac{\chi_s}{\Delta}G(t)P_{\varphi}(0)],
\end{align}
where $A(t)\equiv t-\kappa\int_0^t F(s)ds$, $B(t)\equiv \kappa \int_0^t G(s)ds$, $F(t)\equiv\int_0^t f(s)ds$, $G(t)\equiv\int_0^t g(s)ds$, $Q_{\mathrm{in},\varphi}\equiv \cos\varphi Q_{\mathrm{in}}+\sin\varphi P_{\mathrm{in}}$, $P_{\mathrm{in},\varphi}\equiv \cos\varphi P_{\mathrm{in}}-\sin\varphi Q_{\mathrm{in}}$, and $\varphi^\prime\equiv \varphi-\theta_\lambda$. From Eq.\:\eqref{Signal}, it is evident that when the resonator field starts in a vacuum state, the contrast between the output signals is determined by $2|B(t)\chi_s\Delta^{-1}P_{\mathrm{in},\varphi}|$, which relies on both the input quadratures $Q_{\mathrm{in}}$ and $P_{\mathrm{in}}$, and the local oscillator phase $\varphi$. Notably, in homodyne detection characterized by a local oscillator phase $\varphi$, the signal contrast is proportional to $|\sin(\theta_\alpha-\varphi)|$. In order to maximize the contrast between the output signals, it is necessary for the phase of the displacement amplitude  leads or lags that of the local oscillator by $\pi/2$, i.e., $\theta_\alpha=\varphi\pm \pi/2$. 

To proceed, one can choose the input field as a displaced squeezed vacuum state $|\boldsymbol{\alpha},\boldsymbol{\xi}\rangle \equiv D(\boldsymbol{\alpha})S(\boldsymbol{\xi})|0\rangle$, where $D(\boldsymbol{\alpha})\equiv\exp\int dk (\alpha_k b_k^\dagger-\alpha_k^*b_k)$ is a continuous displacement operator, and $S(\boldsymbol{\xi})\equiv \exp\frac{1}{2}\int dk(\xi_k^* b_k^2-\xi_k b_k^{\dagger 2})$ is a continuous squeezing operator \cite{fedorov2016displacement, mandel1995optical}. Notably, the expectation values of the annihilation and creation operators coincide with those of coherent states, i.e., $\langle b_{\mathrm{in}}\rangle=\alpha(t)$ and $\langle b^\dagger_{\mathrm{in}}\rangle=\alpha^*(t)$. Alternatively, one can also use the squeezed coherent state $|\boldsymbol{\xi},\boldsymbol{\gamma}\rangle\equiv S(\boldsymbol{\xi})D(\boldsymbol{\gamma})|0\rangle$ as an input field. It can be regarded as a displaced squeezed vacuum state with identical squeezing parameters, yet characterized by distinct displacement amplitudes $\alpha_k=\gamma_k\cosh r_k-\gamma_k^*\sinh r_k e^{i\theta_{\xi_k}}$. Conversely, the displaced squeezed vacuum state can be considered as a squeezed coherent state with displacement amplitudes $\gamma_k=\alpha_k\cosh r_k+\alpha_k^*\sinh r_ke^{i\theta_{\xi_k}}$. In this state, the total photon number remains the same, i.e., $\int \langle b^\dagger_{\mathrm{in}}b_{\mathrm{in}}\rangle dt= |\alpha|^2+\sinh^2 r$.

\begin{figure}[tbp]
	\subfloat[$\kappa=\chi_s$\label{sfig:2a}]{%
	\includegraphics[width=0.484\columnwidth]{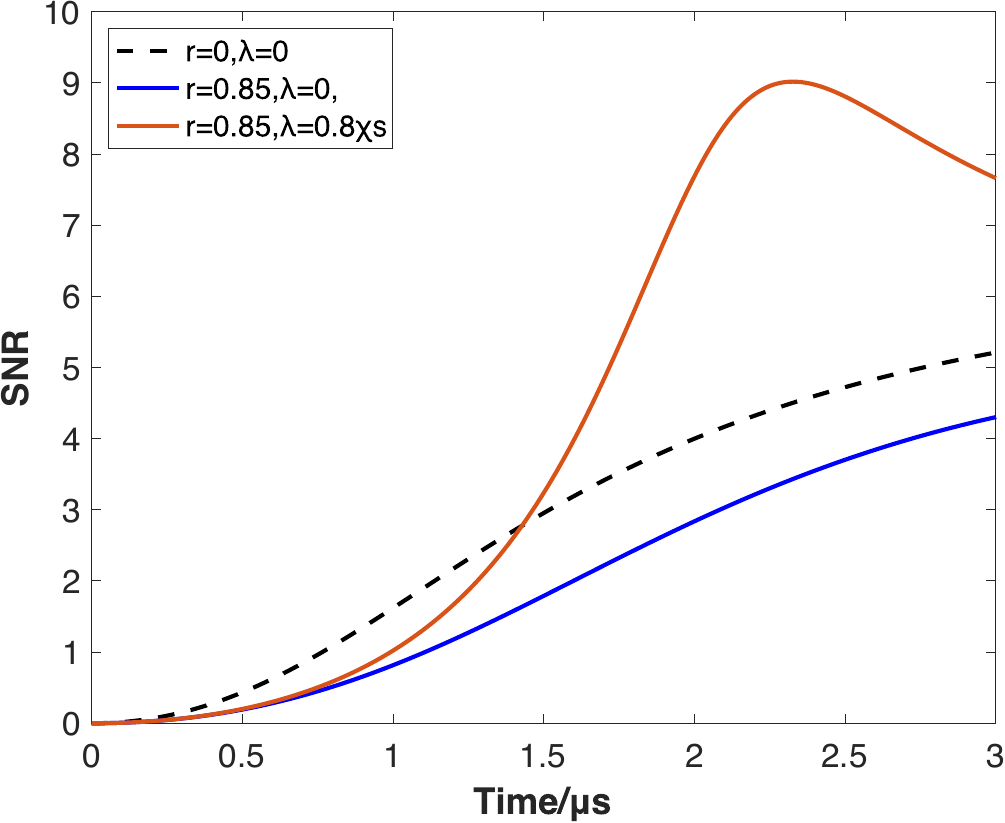}%
	}\hfill
	 \subfloat[$\kappa=\chi_s$\label{sfig:2b}]{%
	\includegraphics[width=0.515\columnwidth]{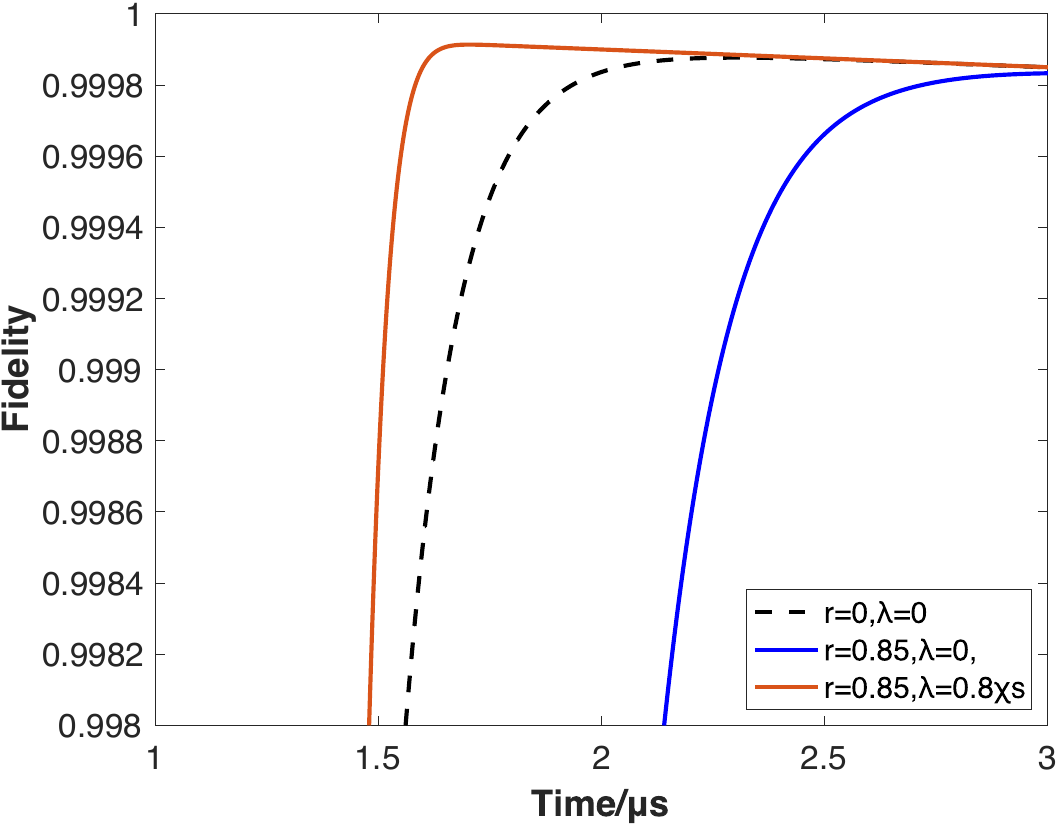}%
	 }\hfill
	\subfloat[$\kappa=2\chi_s$\label{sfig:2c}]{%
 	 \includegraphics[width=0.484\columnwidth]{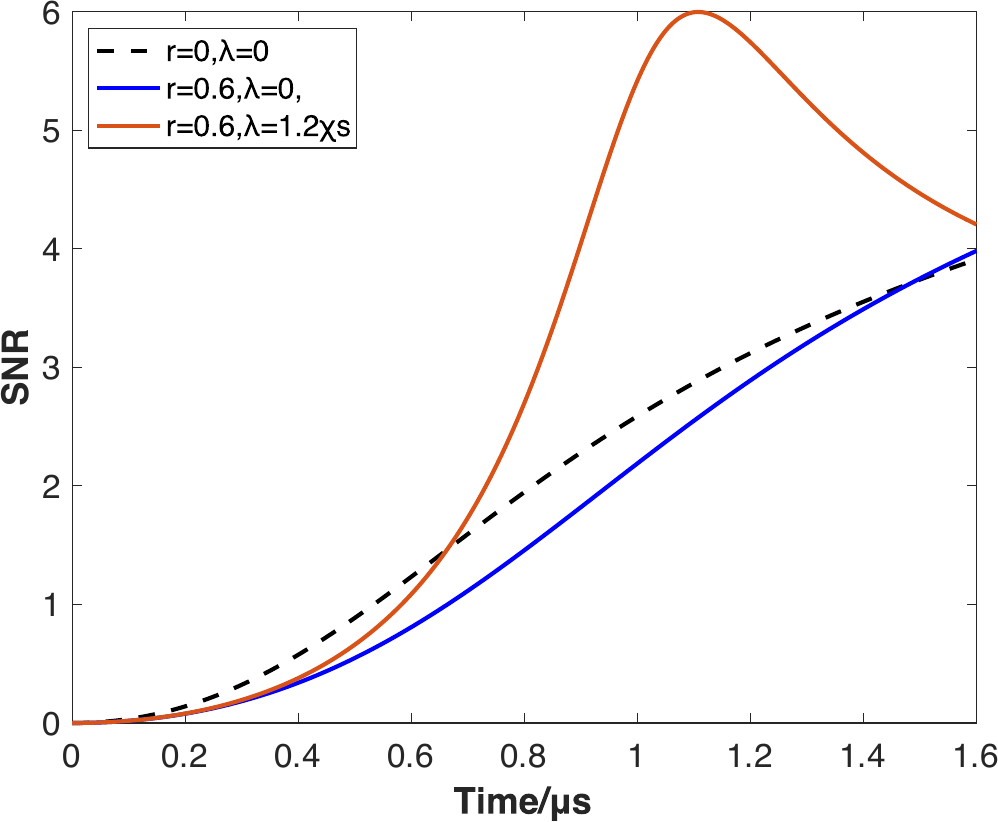}%
	}\hfill
	\subfloat[$\kappa=2\chi_s$\label{sfig:2d}]{%
 	 \includegraphics[width=0.515\columnwidth]{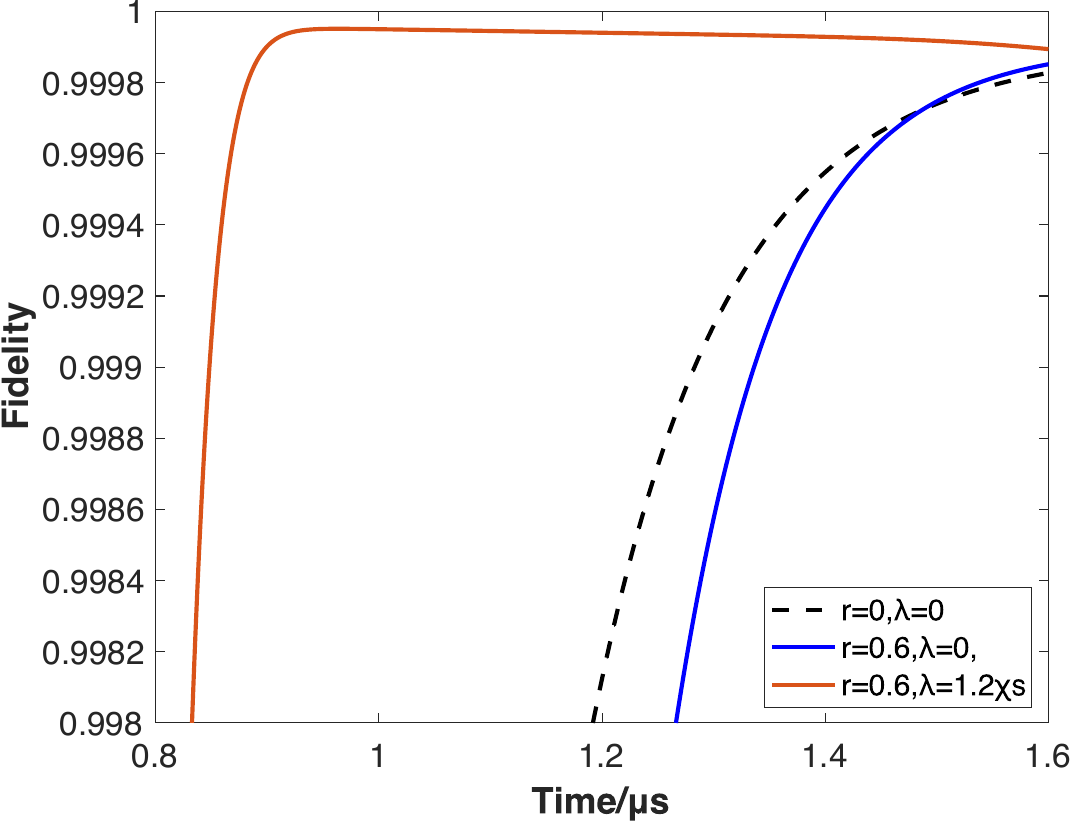}%
	}
\caption{Schematics of the signal-to-noise ratio ($\mbox{SNR}$) and the single-shot readout fidelity with respect to different leakage rates $\kappa$, squeezing parameters $r$, and nonlinear strengths $\lambda$. Here, the dispersive coupling strength, the amplitude of the input displaced squeezed vacuum state, and the spin $T_1$ relaxation time are: $\chi_s=2\pi\times 0.15\:\mbox{MHz}$, $|\alpha|=\sqrt{30}$, and $T_1=10\:\mbox{ms}$. The black dashed lines represent the coherent state inputs.}
\label{fig:Fig2}
\end{figure}

To simplify the preceding discussion, we will first examine the scenario with $\lambda_y=0$ and $\theta_\xi=\pi$. Particularly, for the case with $\varphi=\pi/2$, the standard deviations of the output signals for the two qubit states can be written as (see Appx.\:\ref{A})
\begin{equation}\label{Short}
(\Delta \mathscr{M}^\pm(t))^2=\frac{e^{-2r}}{2}A^{\prime 2}(t)+\frac{e^{2r}}{2}B^{\prime 2}(t)+\frac{\kappa}{2}(F^{\prime 2}(t)+G^{\prime 2}(t)),
\end{equation}
where the functions $A'(t)$, $B'(t)$, $F'(t)$, $G'(t)$ are determined by 
\begin{subequations}
\begin{gather}\label{A1}
A^\prime(t) \equiv A(t)+\frac{\lambda_x}{\Delta}B(t),B^\prime(t)\equiv \frac{\chi_s}{\Delta}B(t),\\
F^\prime(t) \equiv F(t)-\frac{\lambda_x}{\Delta}G(t),G^\prime(t)\equiv \frac{\chi_s}{\Delta}G(t).
\end{gather}
\end{subequations}
Referring to Eq.\:\eqref{Short}, one can reasonably infer that this approach holds the potential to enhance the SNR for short read out times. Given that $A(t)$ behaves as $t-\frac{1}{2}\kappa t^2$ and $B(t)$ behaves as $\frac{1}{6}\kappa\chi_st^3$, it is evident that $B(t)$ is significantly smaller than $A(t)$ for short readout times. It follows that, during a fast dispersive readout, the primary contribution to the noise $\Delta \mathscr{M}^\pm(t)$, i.e., the first term of Eq.\:\eqref{Short}, would be largely suppressed by the incorporation of external squeezing. Moreover, it is evident from Eqs.\:\eqref{Short} and \eqref{A1} that, the primary contribution to noise $\Delta \mathscr{M}^\pm(t)$, specifically the component proportional to $A^\prime(t)$, diminishes when utilizing an appropriate negative value for $\lambda_x$, or equivalently setting $\theta_\lambda=\pi$. This observation can be validated through a more comprehensive analysis in subsequent sections. 

In contrast, for the case with $\varphi=0$, the standard deviations of the output signals for the two qubit states can be written as
\begin{equation}\label{NewNoise}
(\Delta \mathscr{M}^\pm(t))^2=\frac{e^{2r}}{2}A^{\prime 2}(t)+\frac{e^{-2r}}{2}B^{\prime 2}(t)+\frac{\kappa}{2}(F^{\prime 2}(t)+G^{\prime 2}(t)),
\end{equation}
where the functions $A'(t)$, $B'(t)$, $F'(t)$, $G'(t)$ are determined by 
\begin{subequations}
\begin{gather}\label{A2}
A^\prime(t) \equiv A(t)-\frac{\lambda_x}{\Delta}B(t),B^\prime(t)\equiv \frac{\chi_s}{\Delta}B(t),\\
F^\prime(t) \equiv F(t)+\frac{\lambda_x}{\Delta}G(t),G^\prime(t)\equiv \frac{\chi_s}{\Delta}G(t).
\end{gather}
\end{subequations}
Upon examining Eq.\:\eqref{NewNoise}, one may notice that, unlike in the previous case, the primary factor influencing the noise $\Delta \mathscr{M}^\pm(t)$, i.e., the first term of Eq.\:\eqref{NewNoise}, is enhanced through squeezing rather than diminished. This generally makes this approach less favorable. However, a significant SNR improvement is still possible when $A^\prime(t)$ reaches zero at a specific time. From Eq.\:\eqref{A2}, one can see that the condition for vanishing $A^\prime(t)$ can be met by choosing an appropriate positive value for $\lambda_x$ or, equivalently, setting $\theta_\lambda=0$. This observation can also be validated through a more comprehensive analysis in subsequent sections. 

\begin{figure}[tbp]
	\subfloat[$r=0.6$\label{sfig:3a}]{%
	\includegraphics[width=0.498\columnwidth]{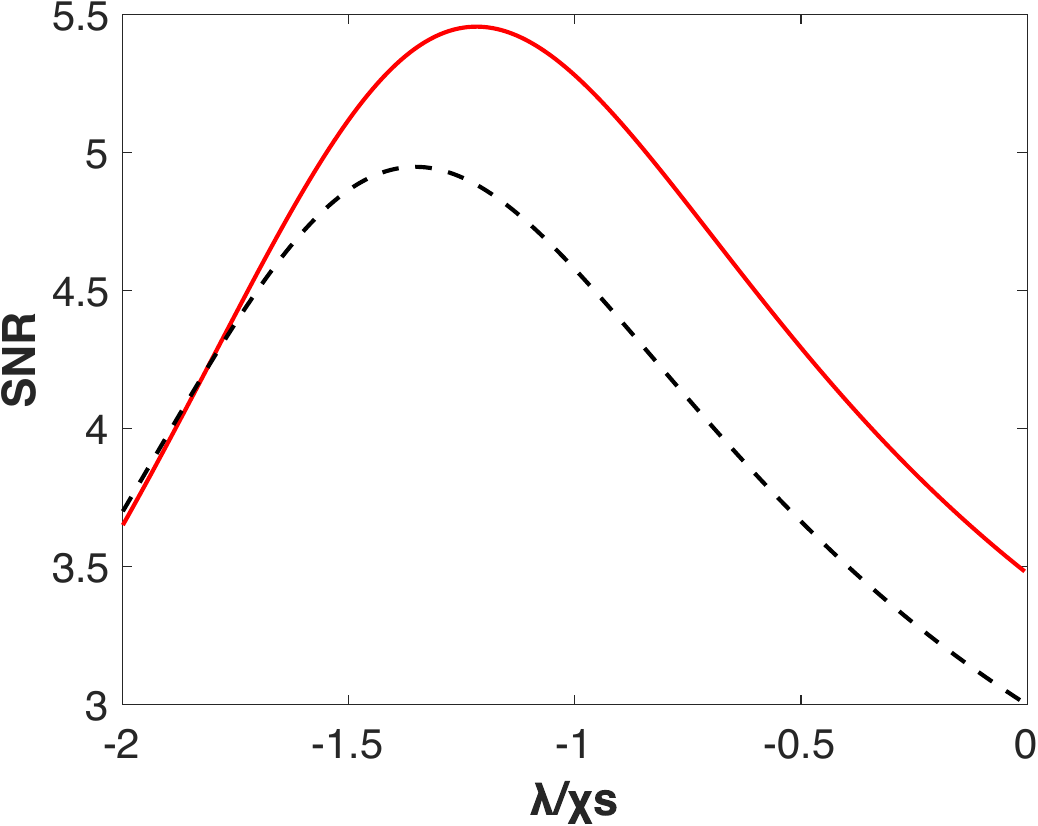}%
	}\hfill
	\subfloat[$\lambda=-1.2\chi_s$\label{sfig:3b}]{%
 	 \includegraphics[width=0.498\columnwidth]{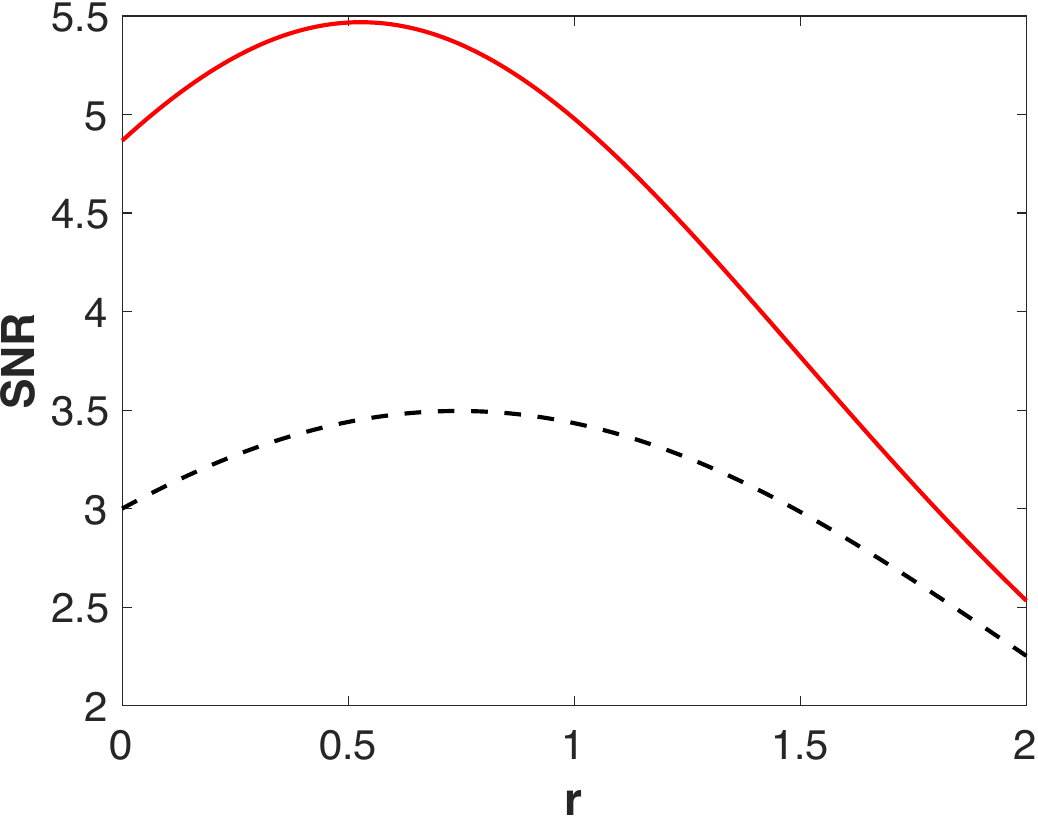}%
	}
\caption{The signal-to-noise ratio (SNR) with respect to different nonlinear strengths and squeezing parameters at a fixed measurement time $t\approx 0.714\:\mu s$. Here, $|\alpha|=10$, $\kappa=2\chi_s$, and $\chi_s=2\pi\times 0.15\:\mbox{MHz}$. The black dashed lines in Figs.\:\ref{sfig:3a} and \ref{sfig:3b} correspond to the cases where squeezing and resonator nonlinearity are absent, respectively.}
\label{fig:Fig3}
\end{figure}

\subsection{Readout fidelity}
When describing the ability to distinguish qubit states, the concepts of signal-to-noise ratio (SNR) and single-shot readout fidelity are intricately interconnected. In the context of an ideal qubit with an infinite longitudinal relaxation time, i.e., $T_1\rightarrow\infty$, the fidelity of a single-shot readout relies exclusively on the SNR of the output signal. Specifically, when the standard derivations of the time-integrated output quadratures are equal ($\Delta \mathscr{M}^+(t)=\Delta \mathscr{M}^-(t)$), the single-shot readout fidelity $\mathscr{F}(t)$ is exclusively determined by the SNR, given by 
\begin{equation}\label{SNRtoF}
\mathscr{F}(t) = \erf\left(\frac{\mbox{SNR}(t)}{\sqrt{2}}\right).
\end{equation}
Here, $\erf z\equiv \frac{2}{\sqrt{\pi}}\int_0^z e^{-s^2}ds$ is the well-known error function \cite{abramowitz1972book}. From Eq.\:\eqref{SNRtoF}, it is evident that the single-shot readout fidelity can always be improved by increasing the signal-to-noise ratio, as the error function is a monotonically increasing function. In this context, a unity fidelity output signal corresponds to an infinitely high signal-to-noise ratio. 

In contrast, unlike an ideal qubit, when dealing with a non-ideal qubit with a finite $T_1$ relaxation time, the non-ideal qubit, initially prepared in the excited state, will decay in a later time. Hence, it becomes necessary to consider the qubit's transition from the excited to the ground state. Consequently, the single-shot readout fidelity will also depend on the qubit's $T_1$ relaxation time. In such a scenario, the single-shot readout fidelity at time $t\ll T_1$ is simply modified as (see Appx.\:\ref{B})
\begin{equation}\label{SNRtoF2}
\mathscr{F}(t) = \exp\left(-\frac{t}{2T_1}\right)\erf\left(\frac{\mbox{SNR}(t)}{\sqrt{2}}\right).
\end{equation}
This modified single-shot readout fidelity accounts for both the relaxation process of the qubit and the impact of noise on the readout. Notably, the primary factor affecting the single-shot readout fidelity is the qubit's $T_1$ relaxation time. This is due to the fact that the dispersive Hamiltonian Eq.\:\eqref{Dispersive} exclusively relates to the population difference between the two qubit states. It's worth noting that the $T_1$ relaxation time of quantum dot spin qubits is typically long, ranging from a few milliseconds to a few seconds, whereas the measurement time used here, which ranges from a few hundred nanoseconds to a few microseconds, is relatively short compared to the $T_1$ relaxation time. In other words, $t/T_1$ is at most $10^{-3}$, which validates our assumption used to derive Eq.\:\eqref{SNRtoF2}. Furthermore, as one can discern from Eq.\:\eqref{SNRtoF2} that, when the condition $t/T_1\ll 1$ is fulfilled, an SNR of approximately 2.6 ensures a fidelity of 99\%, while an SNR of about 3.9 guarantees a fidelity of 99.99\%. In this regard, for our purpose, a sufficiently high SNR with a shorter measurement time is more preferable than an extremely high SNR with a longer measurement time.  

\section{The Results}
\subsection{The phase choices with $\theta_\xi=\pi$, $\theta_\lambda=\pi$ and $\varphi=\pi/2$}\label{IVa}
Fig.\:\ref{sfig:1a} shows an apparent SNR enhancement in sub-microsecond time regimes. This enhancement is achieved by using current technology with $\kappa=\chi_s=2\pi\times 0.15\:\mbox{MHz}$, and employing an input displaced squeezed vacuum state characterized by approximately 30 photons and a modest squeezing parameter of $r = 0.85$ (equivalent to 7.38 dB in decibels), along with the phase setting $\theta_\alpha=0$ and $\theta_{\xi}=\pi$, Specifically, there is a notable nearly twofold enhancement in the readout SNR, increasing from less than 1.4 to nearly 2.4 to at 0.9 $\mu s$ when introducing a resonator nonlinearity of $\lambda=-0.8\chi_s$, in contrast to the scenario without squeezing and nonlinearity. Fig.\:\ref{sfig:1b} demonstrates a universal improvement in readout fidelity achieved by combining the displaced squeezed vacuum state with resonator nonlinearity in the sub-microsecond regime. Specifically, when $T_1=10\:\mbox{ms}$, one can observe a remarkable readout fidelity of 98\% at 0.9 $\mu s$, in contrast to the scenario without squeezing and nonlinearly, which only achieves the fidelity of 82\% at the same readout time. 

Fig.\:\ref{sfig:1c} reveals that a slight increase in the leakage rate to $\kappa=2\chi_s$ results in an almost twofold enhancement in the readout SNR at approximately $0.6\:\mu s$, as compared to the scenario without squeezing and nonlinearity. The SNR increases from just above 1.2 to around 2.3. This improvement is achieved through the utilization of a displaced squeezed vacuum state with approximately 30 photons and a mild squeezing parameter of $r=0.6$ (equivalent to 5.21 dB in decibels). Fig.\:\ref{sfig:1d} illustrates that, when $T_1=10\:\mbox{ms}$, the selected chosen squeezed vacuum state, in conjunction with a microwave resonator characterized by a nonlinear strength of $\lambda=-1.2\chi_s$, attains a notable readout fidelity of 97.8\% at around $0.6\:\mu s$, in contrast to the modest 78.2\% achieved without squeezing or 88\% with only squeezing but lacking resonator nonlinearly.

\begin{figure}[tbp]
	\subfloat[$t=0.1\:\mu\mbox{s}$\label{sfig:4a}]{%
	\includegraphics[width=0.498\columnwidth]{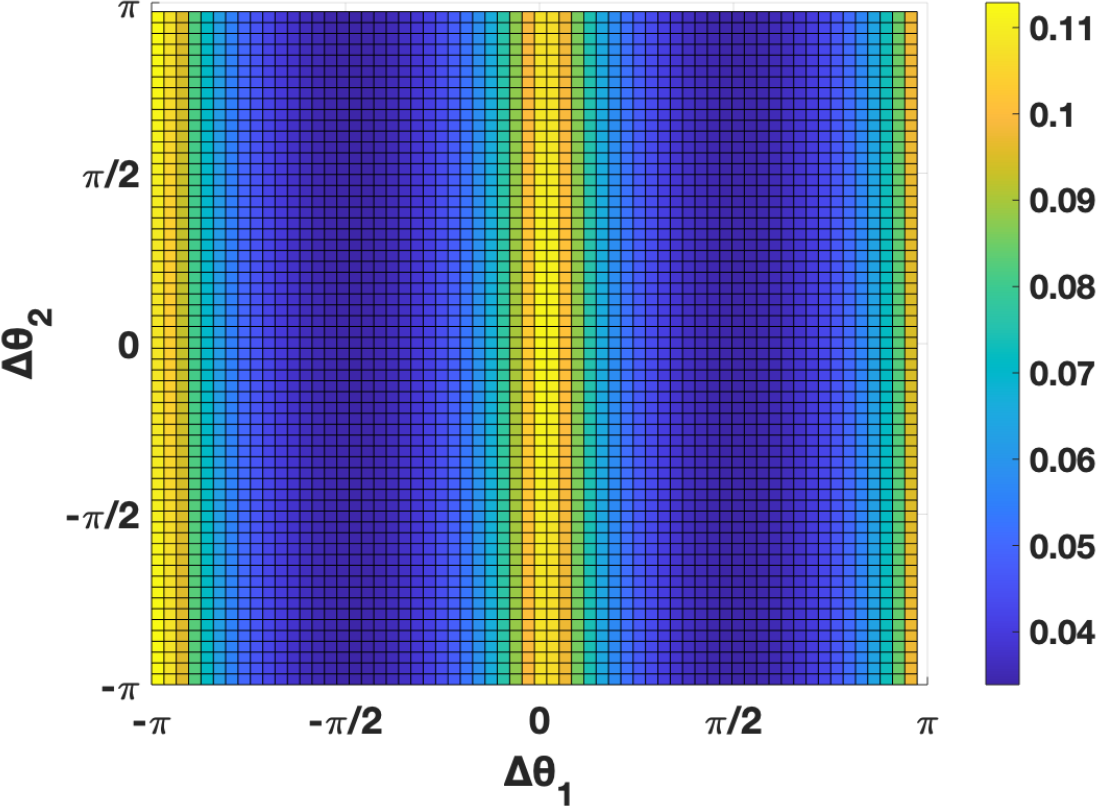}%
	}\hfill
	\subfloat[$t=0.35\:\mu\mbox{s}$\label{sfig:4b}]{%
 	 \includegraphics[width=0.498\columnwidth]{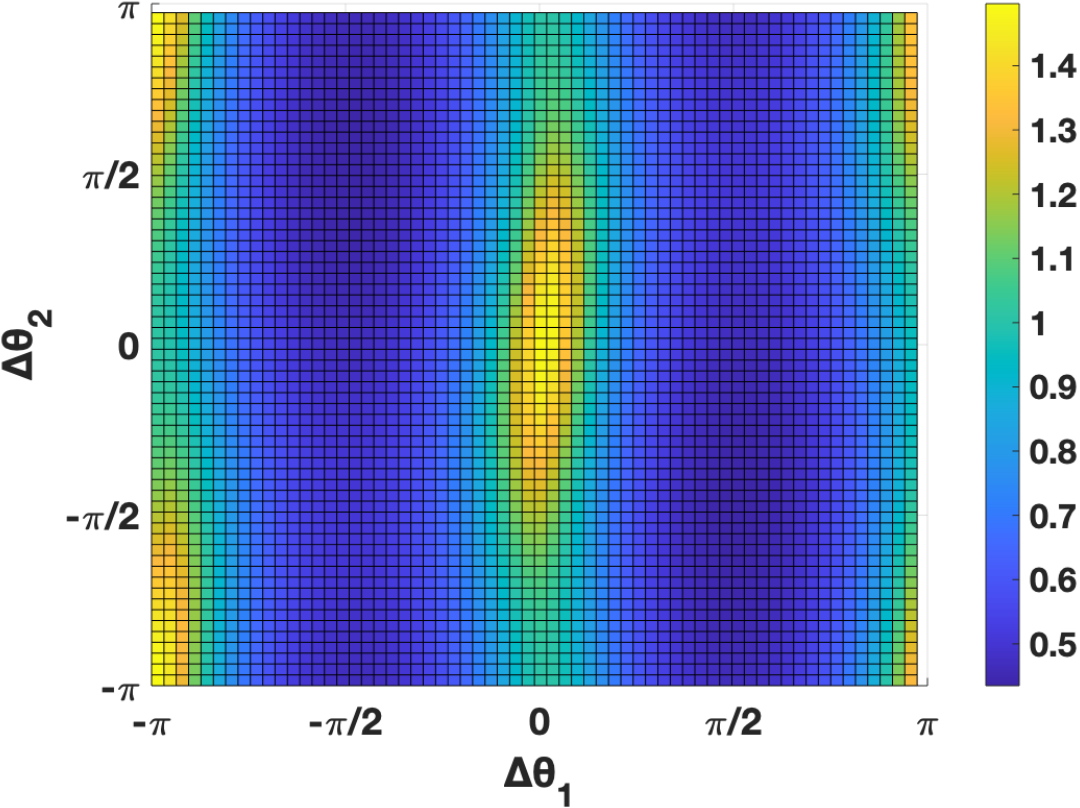}%
	}\hfill
	\subfloat[$t=0.6\:\mu\mbox{s}$\label{sfig:4c}]{%
 	 \includegraphics[width=0.498\columnwidth]{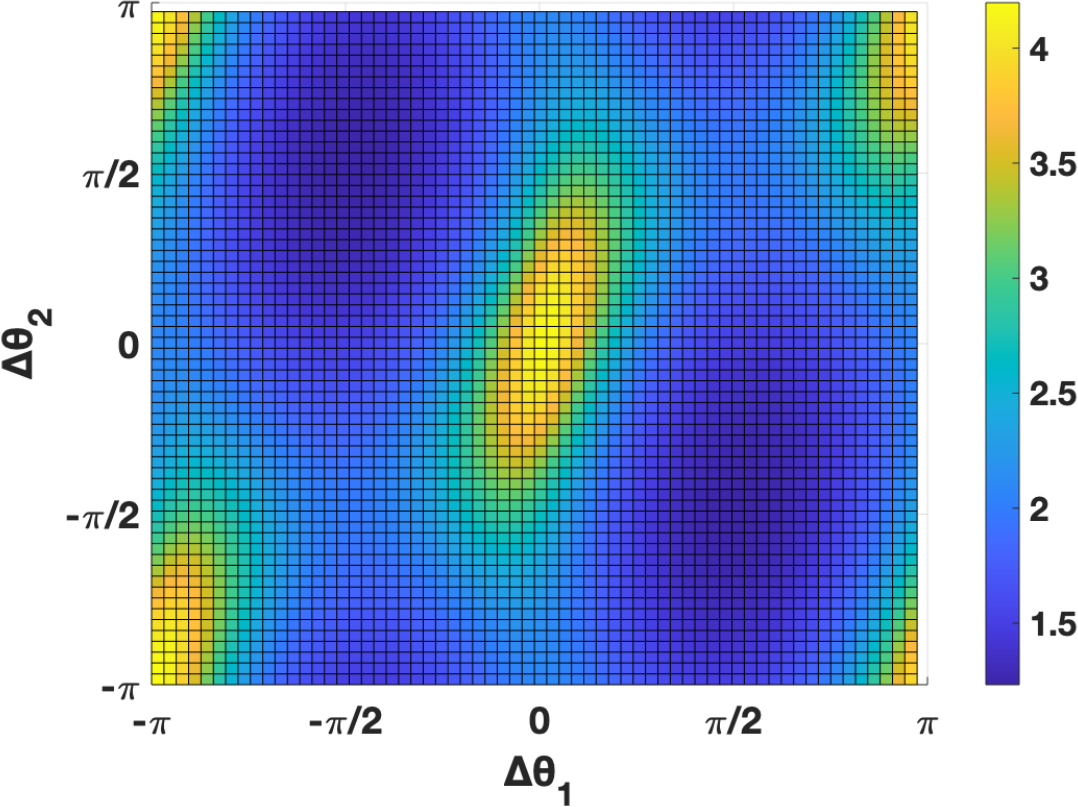}%
	}\hfill
	\subfloat[$t=0.85\:\mu\mbox{s}$\label{sfig:4d}]{%
 	 \includegraphics[width=0.498\columnwidth]{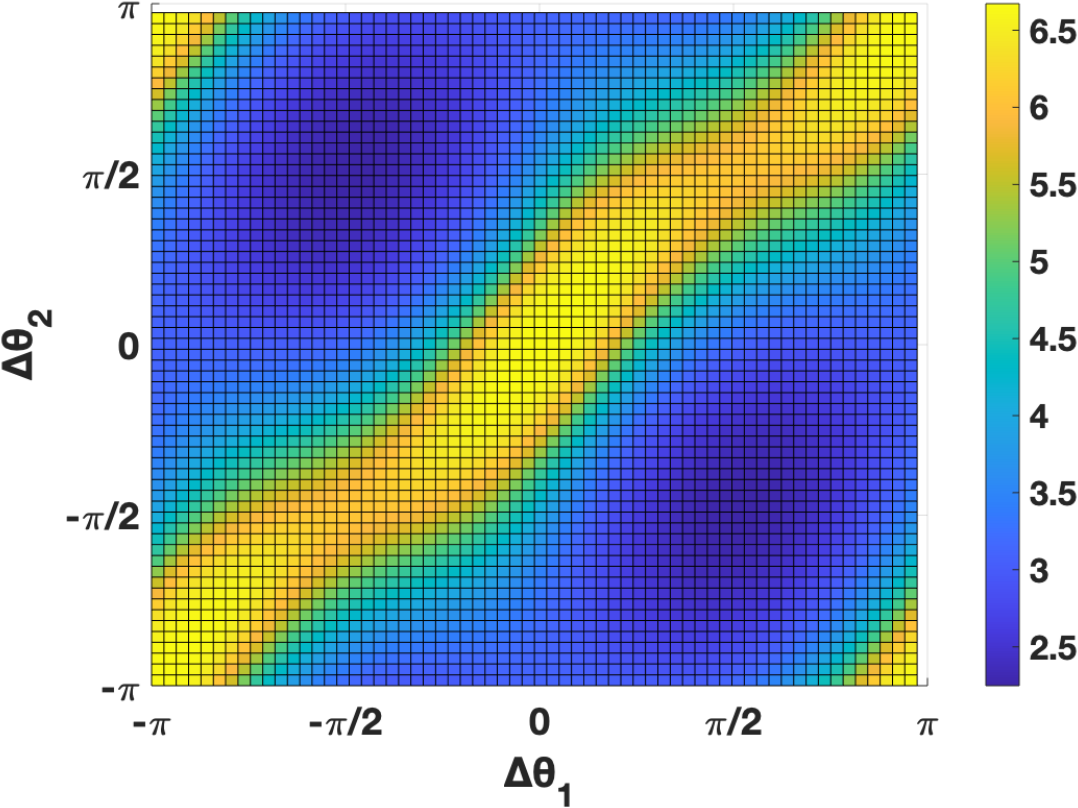}%
	}\hfill
	\subfloat[$t=1.1\:\mu\mbox{s}$\label{sfig:4e}]{%
 	 \includegraphics[width=0.498\columnwidth]{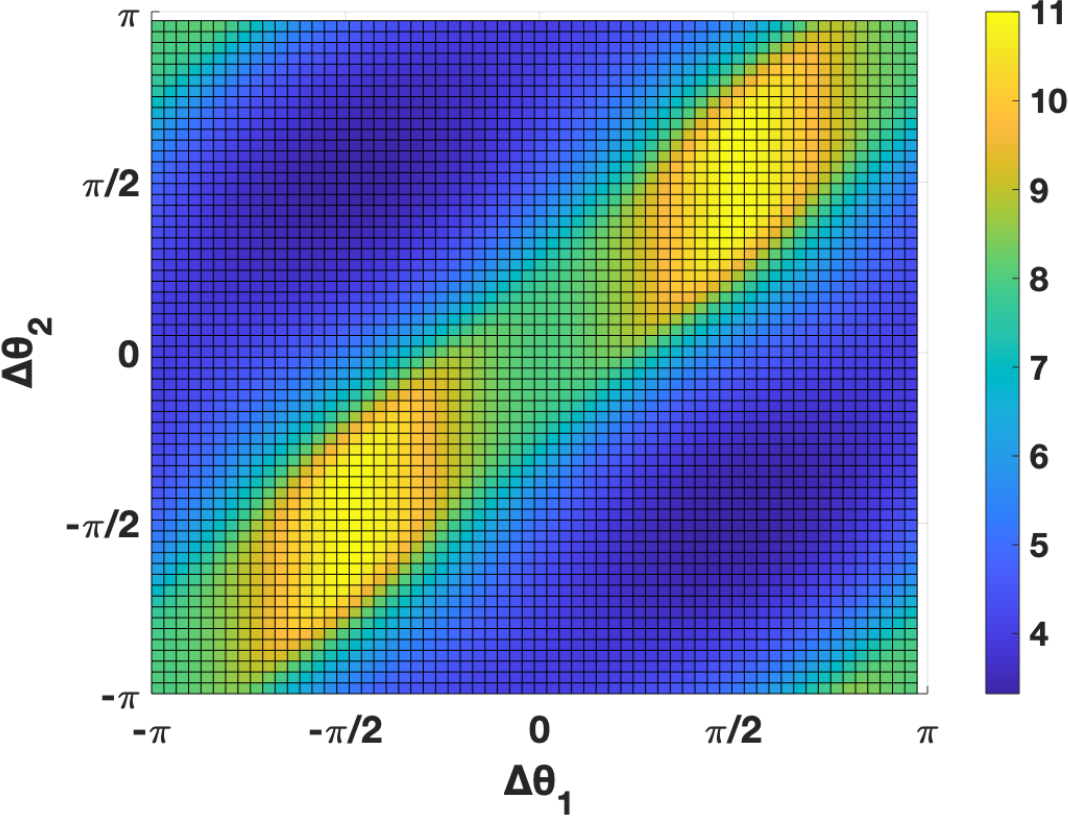}%
	}\hfill
	\subfloat[$t=1.6\:\mu\mbox{s}$\label{sfig:4g}]{%
 	 \includegraphics[width=0.498\columnwidth]{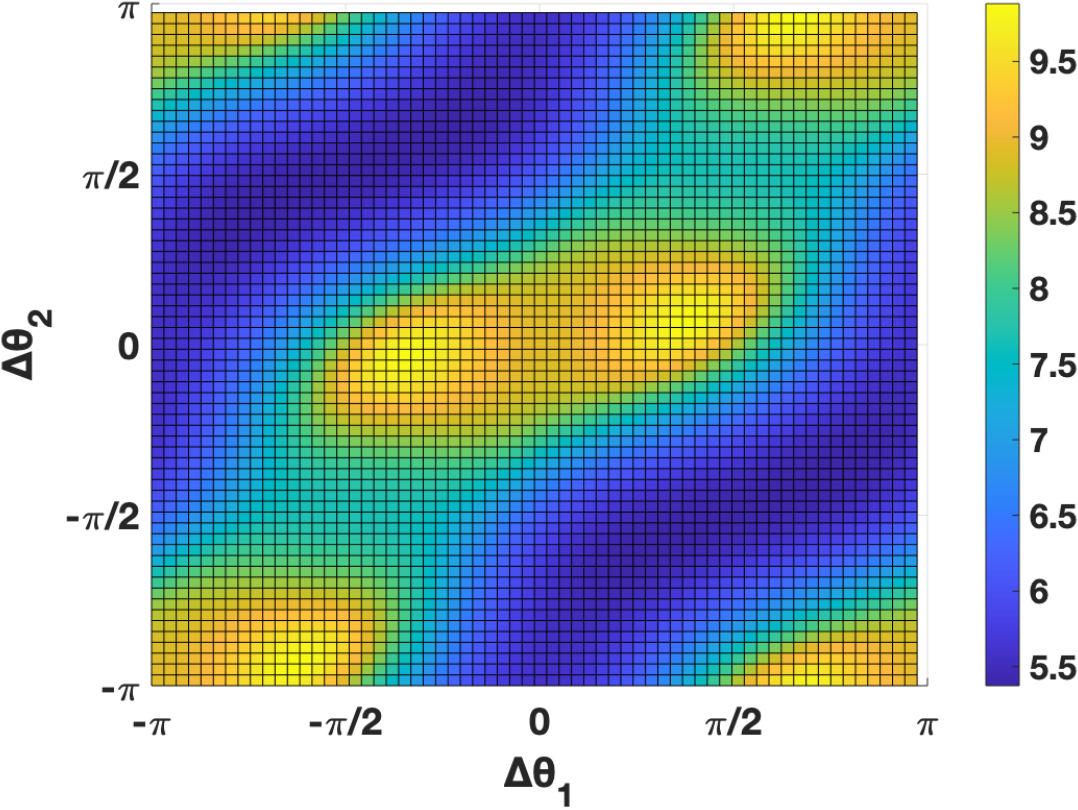}%
	}\hfill
	\subfloat[$t=2.1\:\mu\mbox{s}$\label{sfig:4i}]{%
 	 \includegraphics[width=0.498\columnwidth]{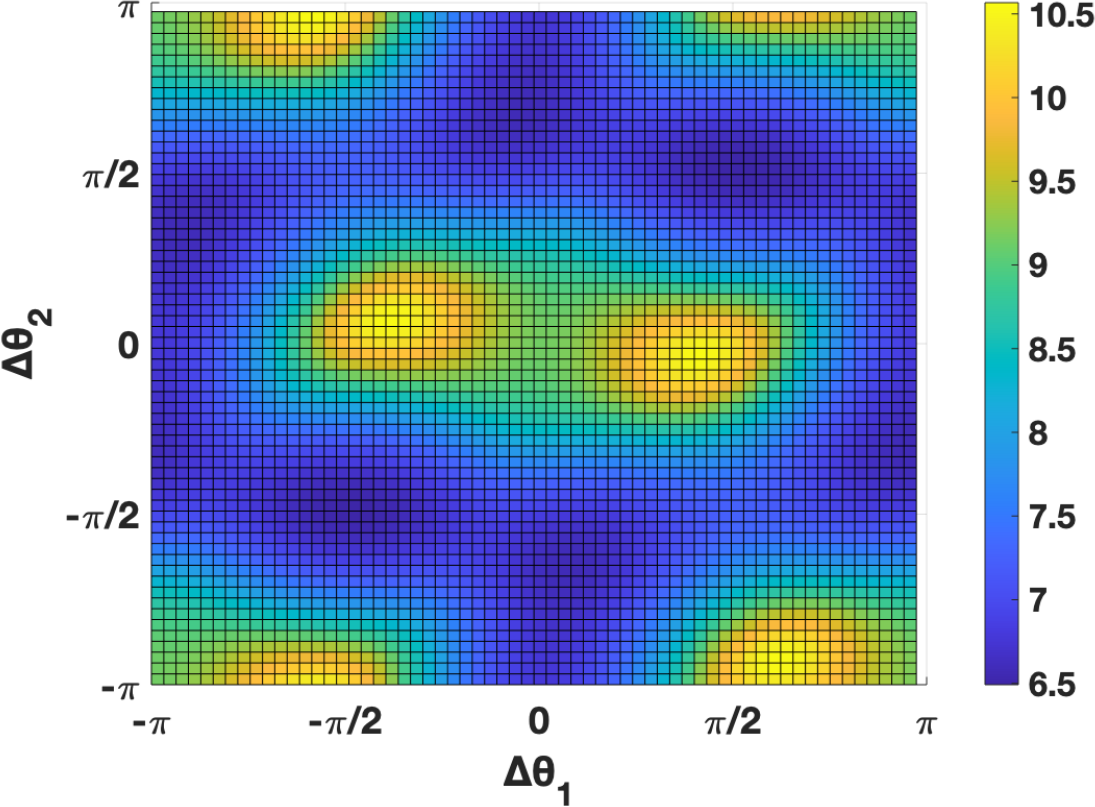}%
  	}\hfill
	\subfloat[$t=5\:\mu\mbox{s}$\label{sfig:4m}]{%
 	 \includegraphics[width=0.498\columnwidth]{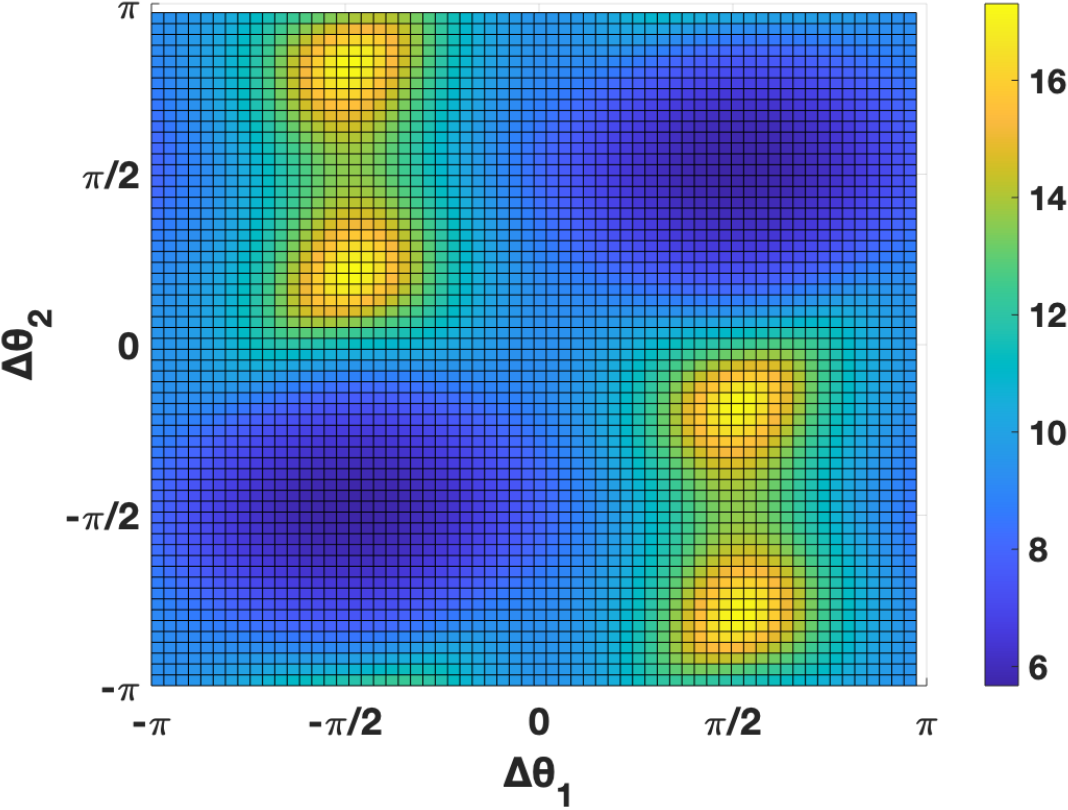}%
	}
\caption{The signal-to-noise ratio (SNR) with respect to different phase choices $\Delta\theta_1$ and $\Delta\theta_2$, where $\Delta\theta_1\equiv\theta_\xi/2-\varphi$, $\Delta\theta_2\equiv \theta_\xi/2+\varphi^\prime$, $\varphi^\prime \equiv \varphi-\theta_\lambda$. Here, $|\alpha|=10$, $\kappa=2\chi_s$, $\chi_s=2\pi\times 0.15\:\mbox{MHz}$, $r=0.6$, and $|\lambda|=1.2\chi_s$.}
\label{fig:Fig4}
\end{figure}

\subsection{The phase choices with $\theta_\xi=\pi$, $\theta_\lambda=0$ and $\varphi=0$}\label{IVb}
Unlike the previous cases, the combination of an input displaced squeezed state and a resonator nonlinearity, along with the specific phase choices $\theta_\xi=\pi$, $\theta_\lambda=0$ and $\varphi=0$ results in an apparent peak in the output SNR measured at a specific time in later times. For example, as illustrated in Fig.\:\ref{sfig:2a}, when the leakage rate $\kappa$ is equal to the dispersive coupling strength $\chi_s$, where $\chi_s=2\pi\times 0.15\:\mbox{MHz}$ and $T_1=10\:\mbox{ms}$, there exists a notable peak in the readout SNR, reaching approximately 9 exactly at around $2.3\:\mu s$. This is achieved by using a displaced squeezed vacuum state containing approximately 30 photons and a modest squeezing parameter of $r = 0.85$ (equivalent to 7.38 dB in decibels), along with the introduction of a resonator nonlinearity with a value of $\lambda=0.8\chi_s$. As shown in Fig.\:\ref{sfig:2b}, this phase choice yields a fidelity of 99.99\% at around $1.6\:\mu s$, in contrast to the case without squeezing and nonlinearly, which only achieves the fidelity of 99.85\% at the same readout time.

As another example, as shown in Fig.\:\ref{sfig:2c}, when the leakage rate $\kappa$ is twice the dispersive coupling strength $\chi_s$, where $\chi_s=2\pi\times 0.15\:\mbox{MHz}$ and $T_1=10\:\mbox{ms}$, a distinct peak becomes apparent the readout SNR, reaching approximately 6 exactly at around $1.1\:\mu s$. This is attained by using a displaced squeezed vacuum state containing approximately 30 photons and a modest squeezing parameter of $r=0.6$ (equivalent to 5.21 dB in decibels), along with the introduction of a resonator nonlinearity with a value of $\lambda=1.2\chi_s$. As depicted in Fig.\:\ref{sfig:2d}, the peak in the SNR leads to a substantial enhancement in the readout fidelity, approaching nearly 99.99\%, at approximately $0.9\:\mu s$, in contrast to the case without squeezing and nonlinearly, which only achieves the fidelity of 97.73\% at the same readout time.

Comparing the two scenarios discussed above, it is evident that one should adhere to the former scenario, characterized by the phase choice with $\theta_\lambda=\pi$ and $\varphi=\pi/2$, if the primary objective is to achieve a relatively high SNR and fidelity in an ultrafast readout time. In contrast, one may employ the latter scenario, characterized by the phase choice with $\theta_\lambda=\varphi=0$, if the primary objective is to achieve an ultrahigh SNR and fidelity, albeit at a relatively later readout time. In the latter scenario, the precise time when the SNR attaining its maximum is determined by $t_c=\tau_c/\Delta$, where $\tau_c$ is a dimensionless constant depending only on the leakage rate $\kappa$, the nonlinear strength $\lambda$ and the dispersive coupling strength $\chi_s$ (see Appx.\:\ref{A}).

\subsection{Optimization of the squeezing parameter $r$ and the nonlinear strength $\lambda$}
As our primary goal in this study is to attain a relatively high SNR and fidelity in the shortest possible readout time, we focus on optimizing parameters in the former scenario with $\theta_\lambda=\pi$ and $\varphi=\pi/2$. To maximally enhance the contrast between the output signals, we set the phase of the displacement amplitude to $\theta_\alpha=\varphi\pm \pi/2$, resulting in values of $0$ or $\pi$ in the current scenario. Additionally, we set the leakage rate $\kappa$ to be twice the dispersive coupling strength $\chi_s$, as this choice yields higher SNR and fidelity for all different phase settings compared to the case where $\kappa$ equals $\chi_s$. 

As depicted in Fig.\:\ref{sfig:3a}, with the squeezing parameter $r$ fixed at $0.6$, it is evident that the peak of the SNR corresponds to a negative nonlinear strength of value $\lambda\approx -1.218\chi_s$, while the SNR drops for both positive nonlinear strengths as well as large negative nonlinear strengths. Notably, the black dashed line illustrates that, even in the absence of squeezing, the resonator nonlinearity can still enhance the SNR when employing a slightly more negative nonlinear strength. 

In Fig.\:\ref{sfig:3b}, one can observe that with a constant nonlinear strength of $\lambda\approx -1.2\chi_s$, the SNR reaches its peak at a mild squeezing level, characterized by a squeezing parameter of about $r\approx 0.53$, while the SNR decreases for both minimal and excessive squeezing levels. Notably, the black dashed line illustrates that, even in the absence of resonator nonlinearity, a slightly larger squeezing parameter can still enhance the SNR, although the peak is lower than that with resonator nonlinearity.

\begin{table}
\centering
\begin{tabular}{|p{2.2cm}|p{1.2cm}||p{0.9cm}|p{3.2cm}|}
 \hline
 $t$ & SNR & $\theta_\xi$ & $\theta_\lambda$\\
 \hline
 $\lessapprox 0.1\:\mu\mbox{s}$ & $\ll 1$& $2\varphi$ & arb. \\
 $\approx 0.1\:\mu\mbox{s}\sim 0.6\:\mu\mbox{s}$ &   $\approx 1\sim 4$  & $2\varphi$ & $2\varphi$\\
 $\approx 0.6\:\mu\mbox{s}\sim 1.1\:\mu\mbox{s}$ & $\approx 4 \sim 11 $ & arb. & $2\varphi$\\
 $\approx 1.1\:\mu\mbox{s}$ & $\approx 11$ & $2\varphi+\pi$ & $2\varphi$\\
 $\gtrapprox 5\:\mu\mbox{s}$ & >10 & $2\varphi+\pi$ & $2\varphi+\pi\pm\left(\frac{\pi}{2}-\arcsin\left(\frac{|\lambda|}{2\chi_s}\right)\right)$\\
 \hline
\end{tabular}
\caption{The phase matching conditions across different readout times. Here, the parameters are set as follows: $|\alpha|=10$, $\kappa=2\chi_s$, $\chi_s=2\pi\times 0.15\:\text{MHz}$, $r=0.6$, and $|\lambda|=1.2\chi_s$. Note that the phases $\theta_\xi$ and $\theta_\lambda$ are determined within a range spanning from 0 to $2\pi$.}
\label{T1}
\end{table}

\subsection{The Phase matching conditions}
As previously discussed, the readout SNR depends exclusively on three phases: $\theta_\alpha$, $\Delta\theta_1$, and $\Delta\theta_2$, where $\Delta\theta_1\equiv \theta_\xi/2-\varphi$, $\Delta\theta_2\equiv\theta_\xi/2+\varphi-\theta_\lambda$. To maximize the contrast between the output signals, the phase $\theta_\alpha$ is fixed to be $\varphi+\frac{1}{2}(2l+1)\pi$, where $l$ is an arbitrary integer. Hence, one only needs to explore the phase matching conditions with respect to $\Delta\theta_1$ and $\Delta\theta_2$. Without loss of generality, in the following, we will fix $|\alpha|=10$, $\kappa=2\chi_s$, $\chi_s=2\pi\times 0.15\:\mbox{MHz}$, $r=0.6$, and $|\lambda|=1.2\chi_s$.

As depicted in Fig.\:\ref{sfig:4a}, when the readout time is extremely small, corresponding to a significantly low SNR (SNR $\ll 1$), one can deduce from the parallel bright strips along $\Delta\theta_1=m\pi$ that the optimal phase choice should be $\Delta\theta_1\equiv \theta_\xi/2-\varphi=m\pi$, where $m$ denotes an arbitrary integer. This implies that the phase of the squeezed state is determined by the local oscillator phase as $\theta_\xi=2\varphi+2m\pi$. Notice that in this scenario, the choice of the second phase, $\Delta\theta_2\equiv \theta_\xi/2+\varphi-\theta_\lambda$, or the phase of the nonlinear strength, $\theta_\lambda$, has minimal influence on the SNR. 

Fig.\:\ref{sfig:4b} and Fig.\:\ref{sfig:4c} illustrate the transitional scenarios when the readout time is slightly longer (but still in the sub-microsecond regime), characterized by a moderately increased SNR (SNR $\approx 1\sim 4$), wherein the longitudinal bright strips in the SNR diagram gradually transform into bright spots at both the center and the corners. In these scenarios, the phase matching conditions resulting the SNR peaks become $\Delta\theta_1\equiv \theta_\xi/2-\varphi=m\pi$ and $\Delta\theta_2=\theta_\xi/2+\varphi-\theta_\lambda=n\pi$, where $m$ and $n$ are integers such that $m-n$ are even. This implies that the phase of the squeezed state and the phase of the nonlinear strength are determined by the local oscillator phase as $\theta_\xi=2\varphi +2m\pi$ and $\theta_\lambda=2\varphi+(m-n)\pi$. Specifically, when $\varphi=\pi/2$, it precisely corresponds to the special case discussed in Sec.\:\ref{IVa} with $\theta_\xi=\theta_\lambda=2\varphi=\pi$.

Fig.\:\ref{sfig:4d} illustrates the transition as the readout time approaches the microsecond regime, causing the bright spots at the center and corners in the SNR diagram to extend and merge into the diagonal bright strip. This corresponds to the phase matching condition $\Delta\theta_1=\Delta\theta_2+2k\pi$, or equivalently, $\theta_\lambda= 2\varphi+2k\pi$, where $k$ is an arbitrary integer. In this scenario, the choice of squeeze state phase has minimal impact on the SNR.

Fig.\:\ref{sfig:4e} shows the distinctive moment when the SNR attains its local maximum, leading to the bifurcation of the diagonal bright strip into two bright sports along the diagonal line. This corresponds to the phase matching condition $\Delta\theta_1\equiv \theta_\xi/2-\varphi=\frac{1}{2}(2m+1)\pi$ and $\Delta\theta_2\equiv \theta_\xi/2+\varphi-\theta_\lambda=\frac{1}{2}(2n+1)\pi$, where $m$ and $n$ are integers such that $m-n$ are even. This implies that the phase of the squeezed state and the phase of the nonlinear strength are determined by the local oscillator phase as $\theta_\xi=2\varphi +(2m+1)\pi$ and $\theta_\lambda=2\varphi+(m-n)\pi$. Specifically, when $\varphi=0$, it precisely corresponds to the special case discussed in Sec.\:\ref{IVb} with $\theta_\xi=\pi$ and $\theta_\lambda=\varphi=0$.

Fig.\:\ref{sfig:4g} and Fig.\:\ref{sfig:4i} illustrate the transitional scenarios that occur when the readout time is even longer, resulting in the splitting of the bright spots on the diagonal bright strip, and the eventual disappearance of the diagonal bright strip itself. This implies that the phase choice $\Delta\theta_1=\Delta\theta_2+2k\pi$, or equivalently, $\theta_\lambda= 2\varphi+2k\pi$, with $k$ representing an arbitrary integer, becomes less favorable as time progresses. 

Finally, Fig.\:\ref{sfig:4m} shows that when the readout time is sufficiently long, the diagonal bright strip is completely replaced by two distinct pairs of bright spots, situated in both the upper left and the lower right quadrants. In this scenario, the phase matching conditions resulting the SNR peaks are determined by the intersections of the following straight lines
\begin{subequations}
\begin{align}
\Delta\theta_1&\equiv \frac{\theta_\xi}{2}-\varphi=(2m+1)\frac{\pi}{2},\\
\Delta\theta_2&=\Delta\theta_1\pm\left[\frac{\pi}{2}-\arcsin\left(\frac{|\lambda|}{2\chi_s}\right)\right]+(2k+1)\pi,
\end{align}
\end{subequations}
or equivalently
\begin{subequations}
\begin{align}
\Delta\theta_1&\equiv \frac{\theta_\xi}{2}-\varphi=(2m+1)\frac{\pi}{2},\\
\Delta\theta_2&=\left(2k+m+\frac{3}{2}\right)\pi\pm\left[\frac{\pi}{2}-\arcsin\left(\frac{|\lambda|}{2\chi_s}\right)\right],
\end{align}
\end{subequations}
where $k$ and $m$ are arbitrary integers. As such, the phase of the squeezed state and the phase of the nonlinear strength are determined by the local oscillator phase as
\begin{subequations}
\begin{align}
\theta_\xi&=2\varphi+(2m+1)\pi,\\
\theta_\lambda &= 2\varphi-(2k+1)\pi\pm \left[\frac{\pi}{2}-\arcsin\left(\frac{|\lambda|}{2\chi_s}\right)\right].\label{ThetaLambda}
\end{align}
\end{subequations}
As one can see from Table \ref{T1}, it is always beneficial to choose the phase of the squeezed state and the phase of the nonlinear strength as $\theta_\xi=\theta_\lambda=2\varphi$ in the sub-microsecond range, while in the microsecond range, it is better to employ the phase choice $\theta_\xi=2\varphi+\pi$ and $\theta_\lambda=2\varphi$. For a sufficiently long readout time, the optimal squeezed state phase remains the same, but the optimal phase of the nonlinear strength is determined by Eq.\:\eqref{ThetaLambda}.
 
\section{Conclusion and Discussion}
In this study, we suggested a readout scheme that combines displaced squeezed vacuum states and resonator nonlinearity to enhance the readout fidelity and reduce the time required to read a single spin in a semiconductor quantum dot. Our findings reveal that by employing a displaced squeezed vacuum state with specific phase settings (i.e., in homodyne detection characterized by a local oscillator phase $\varphi=\pi/2$, $\theta_\alpha=0$ and $\theta_\xi=\pi)$ and moderate squeezing (a few decibels) can reduce the readout time from microseconds to the sub-microsecond range while maintaining high fidelity. 

For example, with current technology ($\kappa=2\chi_s$,  $\chi_s = 2\pi \times 0.15\:\mbox{MHz}$), using a displaced squeezed vacuum state with 30 photons and a mild squeezing parameter ($r = 0.6$, approximately $5.21$ dB) and a resonator nonlinearity strength $\lambda=-1.2\chi_s$, we can achieve a readout fidelity of 97.8\% within a readout time of about $0.6\:\mu$s, in contrast to the modest 78.2\% achieved without squeezing or 88\% with only squeezing but lacking resonator nonlinearly. Interestingly, a different phase choice such as $\theta_\alpha=\pm\pi/2$, $\theta_\xi=\pi$ and $\varphi=0$ would also yield a substantial enhancement in the readout fidelity, approaching nearly 99.99\%, at approximately $0.9\:\mu s$, with a resonator nonlinearity of $\lambda=1.2\chi_s$, in contrast to the case without squeezing and nonlinearly, which only achieves the fidelity of 97.73\% at the same readout time. In general, when the local oscillator phase $\varphi$ is arbitrary, the phase matching conditions for the maximal signal-to-noise ratio (SNR) at short time are $\theta_\alpha=\varphi\pm\pi/2$, $\theta_\xi=2\varphi$, and $\theta_\lambda=2\varphi$, while those for longer time are $\theta_\alpha=\varphi\pm\pi/2$, $\theta_\xi=2\varphi+\pi$, and $\theta_\lambda=2\varphi$.

\begin{acknowledgements}
We acknowledge financial support by US ARO via grant W911NF1710257.
\end{acknowledgements}

\begin{appendix}
\section{Solution of Langevin equations for resonator fields with intra-resonator nonlinearity}\label{A}
\subsection{Solution of the Langevin equations}
When both external and internal squeezing are employed, the Langevin equation for the resonator field has the form
\begin{equation}
\dot{X}(t)=
M X(t)
-\sqrt{\kappa}X_{\mathrm{in}},
M\equiv -\frac{\kappa}{2}+\lambda_y\tau_x\mp i \chi_s \tau_y+\lambda_x\tau_z,
\end{equation}
where $\tau_x$, $\tau_y$, and $\tau_z$ are the Pauli matrices acting on the two quadratures $Q(t)$ and $P(t)$, and $\lambda\equiv \lambda_x +i\lambda_y\equiv |\lambda|e^{i\theta_\lambda}$ is the complex intra-resonator nonlinear strength. Particularly, for a time-independent input radiation field, i.e., a continuous-wave (CW) radiation, the Langevin equations for the quadratures of the resonator field are solved by
\begin{gather}\label{MainEq}
X(t) = e^{Mt}X(0)-\sqrt{\kappa}\int_0^tds e^{Ms} X_{\mathrm{in}},
\end{gather}
where $e^{Mt}=e^{-\frac{\kappa}{2}t}(\cosh\Delta t+\sinh\Delta t\boldsymbol{n}\cdot\boldsymbol{\tau})\equiv f(t)+g(t)\boldsymbol{n}\cdot\boldsymbol{\tau}$, and
\begin{equation}
    \Delta^2\equiv \lambda_x^2+\lambda_y^2-\chi_s^2,\boldsymbol{n}\equiv \frac{1}{\Delta}(\lambda_y,\mp i\chi_s,\lambda_x).
\end{equation}
Notably, when $|\lambda|^2<\chi_s^2$, $\Delta$ becomes imaginary. In such a case, one obtains $e^{Mt}=e^{-\frac{\kappa}{2}t}(\cos\Delta' t+\sin\Delta' t\boldsymbol{n}'\cdot\boldsymbol{\tau})$, where $\Delta'\equiv i\Delta$ and $\boldsymbol{n}'=-i\boldsymbol{n}$, such that $\Delta^{'2}=\chi_s^2-|\lambda|^2>0$. Particularly, when $\lambda=0$, one recovers the result $e^{Mt}=e^{-\frac{\kappa}{2}t}(\cos\chi_s t\mp i\sin\chi_st\tau_y)$.

From Eq.\:\eqref{MainEq} the resonator field can be written as
\begin{equation*}
    X(t) = \left[f(t) +g(t)\boldsymbol{n}\cdot\boldsymbol{\tau}\right]X(0) 
    - \sqrt{\kappa}\left[F(t) +G(t)\boldsymbol{n}\cdot\boldsymbol{\tau}\right]X_{\mathrm{in}},
\end{equation*}
where $F(t)\equiv\int_0^t f(s)ds$, and $G(t)\equiv\int_0^t g(s)ds$. Utilizing the input-output relation $X_{\mathrm{out}}=X_{\mathrm{in}}+\sqrt{\kappa}X$, the quadratures of the output radiation field can be expressed as
\begin{equation*}
X_{\mathrm{out}}(t)=[1-\kappa F(t)-\kappa G(t)\boldsymbol{n}\cdot\boldsymbol{\tau}]X_{\mathrm{in}}+\sqrt{\kappa}\left[f(t) +g(t)\boldsymbol{n}\cdot\boldsymbol{\tau}\right]X(0),
\end{equation*}
which immediately yields the time-integrated output quadratures
\begin{subequations}
\begin{align}
\int_0^t &X_{\mathrm{out}}(s)ds = [A(t)-B(t)\boldsymbol{n}\cdot\boldsymbol{\tau}]X_{\mathrm{in}} \nonumber\\
&+\sqrt{\kappa}\left[F(t)+G(t)\boldsymbol{n}\cdot\boldsymbol{\tau}\right]X(0),\\
    A(t)&\equiv t-\kappa \int_0^t F(s)ds,B(t)\equiv \kappa \int_0^t G(s)ds.
\end{align}
\end{subequations}
For instance, for the case when $\lambda_y=0$, one will obtain
\begin{subequations}
\begin{align}
\int_0^t &Q_{\mathrm{out}}(s)ds = \left(A(t)-\frac{\lambda_x}{\Delta}B(t)\right)Q_{\mathrm{in}}\pm \frac{\chi_s}{\Delta}B(t)P_{\mathrm{in}}\nonumber\\
&+\sqrt{\kappa}\left[\left(F(t)+\frac{\lambda_x}{\Delta}G(t)\right)Q(0)\mp \frac{\chi_s}{\Delta}G(t)P(0)\right],\\
\int_0^t &P_{\mathrm{out}}(s)ds = \left(A(t)+\frac{\lambda_x}{\Delta}B(t)\right)P_{\mathrm{in}}\mp \frac{\chi_s}{\Delta}B(t) Q_{\mathrm{in}}\nonumber\\
&+\sqrt{\kappa}\left[\left(F(t)-\frac{\lambda_x}{\Delta}G(t)\right)P(0)\pm \frac{\chi_s}{\Delta}G(t)Q(0)\right],
\end{align}
\end{subequations}
which immediately yields
\begin{align}
&\mathscr{M}^\pm(t)\equiv \cos\varphi \int_0^t Q_{\mathrm{out}}(s)ds + \sin\varphi \int_0^t P_{\mathrm{out}}(s)ds\nonumber\\
&=A(t)Q_{\mathrm{in},\varphi}+\frac{\lambda_x}{\Delta}B(t)Q_{\mathrm{in},\pi-\varphi}\pm \frac{\chi_s}{\Delta}B(t) P_{\mathrm{in},\varphi}\nonumber\\
&+\sqrt{\kappa}[F(t)Q_\varphi(0)-\frac{\lambda_x}{\Delta}G(t)Q_{\pi-\varphi}(0)\mp \frac{\chi_s}{\Delta}G(t) P_{\mathrm{in},\varphi}],
\end{align}
where $Q_{\mathrm{in},\varphi}\equiv \cos\varphi Q_{\mathrm{in}}+\sin\varphi P_{\mathrm{in}}$, $P_{\mathrm{in},\varphi}\equiv \cos\varphi P_{\mathrm{in}}-\sin\varphi Q_{\mathrm{in}}$. Particularly, when $\varphi=\pi/2$, one will obtain 
\begin{equation*}
\mathscr{M}^\pm(t) = A^\prime(t)P_{\mathrm{in}} \mp B^\prime(t)Q_{\mathrm{in}}+\sqrt{\kappa}[F^\prime(t)P(0)\pm G^\prime(t)Q(0)],
\end{equation*}
where
\begin{subequations}
\begin{align}\label{Aprime1}
A^\prime(t) \equiv A(t)+\frac{\lambda_x}{\Delta}B(t)\:\:&\mbox{and}\:\:B^\prime(t)\equiv \frac{\chi_s}{\Delta}B(t),\\
F^\prime(t) \equiv F(t)-\frac{\lambda_x}{\Delta}G(t)\:\:&\mbox{and}\:\:G^\prime(t)\equiv \frac{\chi_s}{\Delta}G(t).\label{Aprime2}
\end{align}
\end{subequations}
Similarly, when $\varphi=0$, one will obtain
\begin{equation*}
\mathscr{M}^\pm(t) = A^\prime(t)Q_{\mathrm{in}} \pm B^\prime(t)P_{\mathrm{in}}+\sqrt{\kappa}[F^\prime(t)Q(0)\mp G^\prime(t)P(0)],
\end{equation*}
where
\begin{subequations}
\begin{align}\label{Bprime1}
A^\prime(t) \equiv A(t)-\frac{\lambda_x}{\Delta}B(t)\:\:&\mbox{and}\:\:B^\prime(t)\equiv \frac{\chi_s}{\Delta}B(t),\\
F^\prime(t) \equiv F(t)+\frac{\lambda_x}{\Delta}G(t)\:\:&\mbox{and}\:\:G^\prime(t)\equiv \frac{\chi_s}{\Delta}G(t).\label{Bprime2}
\end{align}
\end{subequations}

\subsection{Explicit expressions of the functions $A(t)$ and $B(t)$}
In order to derive the explicit expressions of the functions $A(t)$ and $B(t)$, one needs the following integration formulas
\begin{align*}
    \int e^{at}\cosh bt dt &= \frac{e^{at}}{a^2-b^2}(a\cosh bt - b\sinh bt),\\
    \int e^{at}\sinh bt dt &= \frac{e^{at}}{a^2-b^2}(a\sinh bt - b\cosh bt),
\end{align*}
which immediately yields
\begin{subequations}
\begin{align}
    F(t)&=-\frac{1}{\frac{\kappa^2}{4}-\Delta^2}\left(\frac{\kappa}{2}f(t)+\Delta g(t)\right)+\frac{\frac{\kappa}{2}}{\frac{\kappa^2}{4}-\Delta^2},\\
    G(t)&=-\frac{1}{\frac{\kappa^2}{4}-\Delta^2}\left(\frac{\kappa}{2}g(t)+\Delta f(t)\right)+\frac{\Delta}{\frac{\kappa^2}{4}-\Delta^2},\\
    A(t) &= -\frac{\frac{\kappa^2}{4}+\Delta^2}{\frac{\kappa^2}{4}-\Delta^2}t -\frac{\kappa(\frac{\kappa^2}{4}+\Delta^2)[f(t)-1]+\kappa^2\Delta g(t)}{(\frac{\kappa^2}{4}-\Delta^2)^2},\\
    B(t) &= \frac{\kappa\Delta}{\frac{\kappa^2}{4}-\Delta^2}t+\frac{\kappa(\frac{\kappa^2}{4}+\Delta^2)g(t)+\kappa^2\Delta [f(t)-1]}{(\frac{\kappa^2}{4}-\Delta^2)^2}.
\end{align}
\end{subequations}
One may simplify the expressions of the functions $F(t)$, $G(t)$, $A(t)$ and $B(t)$ by introducing an auxiliary phase $\delta$
\begin{equation}
\frac{\kappa}{2}\equiv\sqrt{\frac{\kappa^2}{4}-\Delta^2}\cosh\delta\:\:\mbox{and}\:\:\Delta\equiv \sqrt{\frac{\kappa^2}{4}-\Delta^2}\sinh\delta.
\end{equation}
Clearly, the auxiliary phase $\delta$ is real only when $\frac{1}{4}\kappa^2>\Delta^2$, or equivalently $\frac{1}{4}\kappa^2+\chi_s^2>|\lambda|^2$. From the above, one immediately obtains
\begin{align*}
F(t)&=\frac{\sinh\delta}{\Delta}\left[\cosh\delta-e^{-\kappa/2t}\cosh(\Delta t+\delta)\right],\\
G(t)&=\frac{\sinh\delta}{\Delta}\left[\sinh\delta-e^{-\kappa/2t}\sinh(\Delta t+\delta)\right],\\
A(t)&=\frac{\sinh 2\delta}{\Delta}\left[\cosh 2\delta-e^{-\frac{\kappa}{2}t}\cosh(\Delta t+2\delta)\right]-\cosh 2\delta t,\\
B(t)&=\frac{\sinh 2\delta}{\Delta}\left[e^{-\frac{\kappa}{2}t}\sinh(\Delta t+2\delta)-\sinh2\delta\right]+\sinh 2\delta t.
\end{align*}
To proceed, for the general case when $\varphi$ is arbitrary and $\lambda_y\neq 0$, the time-integrated output quadratures can be written as
\begin{subequations}
\begin{align}
&\int_0^t Q_{\mathrm{out}}(s)ds = \left(A(t)-\frac{\lambda_x}{\Delta}B(t)\right)Q_{\mathrm{in}}-\frac{\lambda_y \mp\chi_s}{\Delta}B(t)P_{\mathrm{in}}\nonumber\\
&+\sqrt{\kappa}\left[\left(F(t)+\frac{\lambda_x}{\Delta}G(t)\right)Q(0)+\frac{\lambda_y \mp\chi_s}{\Delta}G(t)P(0)\right],\\
&\int_0^t P_{\mathrm{out}}(s)ds = \left(A(t)+\frac{\lambda_x}{\Delta}B(t)\right)P_{\mathrm{in}}-\frac{\lambda_y \pm\chi_s}{\Delta}B(t) Q_{\mathrm{in}}\nonumber\\
&+\sqrt{\kappa}\left[\left(F(t)-\frac{\lambda_x}{\Delta}G(t)\right)P(0)+\frac{\lambda_y \pm\chi_s}{\Delta}G(t)Q(0)\right].
\end{align}
\end{subequations}
Then the output signals can be explicitly expressed as
\begin{align}\label{Main}
   &\mathscr{M}^\pm(t)\equiv \cos\varphi \int_0^t Q_{\mathrm{out}}(s)ds + \sin\varphi \int_0^t P_{\mathrm{out}}(s)ds\nonumber\\
   =&A(t)Q_{\mathrm{in},\varphi}+ \frac{|\lambda|}{\Delta}B(t)Q_{\mathrm{in},\pi-\varphi^\prime}\pm\frac{\chi_s}{\Delta}B(t)P_{\mathrm{in},\varphi}\nonumber\\
   +&\sqrt{\kappa}[F(t)Q_{\varphi}(0)-\frac{|\lambda|}{\Delta}G(t)Q_{\pi-\varphi^\prime}(0)\mp\frac{\chi_s}{\Delta}G(t)P_{\varphi}(0)],
\end{align}
where $Q_{\mathrm{in},\varphi}\equiv \cos\varphi Q_{\mathrm{in}}+\sin\varphi P_{\mathrm{in}}$, $P_{\mathrm{in},\varphi}\equiv \cos\varphi P_{\mathrm{in}}-\sin\varphi Q_{\mathrm{in}}$, and $\varphi^\prime\equiv  \varphi-\theta_\lambda$. Eq.\:\eqref{Main} can also be written as
\begin{equation}
\mathscr{M}^\pm(t)\equiv \tilde{A}(t)Q_{\mathrm{in}} + \tilde{B}(t)P_{\mathrm{in}}+\sqrt{\kappa}[\tilde{F}(t)Q(0) +\tilde{G}(t)P(0)],
\end{equation}
where
\begin{subequations}
\begin{align}
\tilde{A}(t)&\equiv A(t)\cos\varphi - B(t)\left(\frac{|\lambda|}{\Delta}\cos\varphi^\prime\pm\frac{\chi_s}{\Delta}\sin\varphi\right),\\
\tilde{B}(t)&\equiv A(t)\sin\varphi +B(t)\left(\frac{|\lambda|}{\Delta}\sin\varphi^\prime\pm\frac{\chi_s}{\Delta}\cos\varphi\right),\\
\tilde{F}(t)&\equiv F(t)\cos\varphi +G(t)\left(\frac{|\lambda|}{\Delta}\cos\varphi^\prime\pm \frac{\chi_s}{\Delta}\sin\varphi\right),\\
\tilde{G}(t)&\equiv F(t)\sin\varphi -G(t)\left(\frac{|\lambda|}{\Delta}\sin\varphi^\prime\pm\frac{\chi_s}{\Delta}\cos\varphi\right). 
\end{align}
\end{subequations}
Now, using the results
\begin{subequations}
\begin{gather}
    (\Delta Q_\mathrm{in})^2=\frac{1}{2}\left(e^{2r}\sin^2\left(\frac{\theta_\xi}{2}\right)+e^{-2r}\cos^2\left(\frac{\theta_\xi}{2}\right)\right),\\
    (\Delta P_\mathrm{in})^2=\frac{1}{2}\left(e^{2r}\cos^2\left(\frac{\theta_\xi}{2}\right)+e^{-2r}\sin^2\left(\frac{\theta_\xi}{2}\right)\right),\\
    \cov(Q_\mathrm{in}P_\mathrm{in})=-\frac{1}{2}\sinh 2r\sin\theta_\xi+\frac{i}{2},\\
    \cov(P_\mathrm{in}Q_\mathrm{in})=-\frac{1}{2}\sinh 2r\sin\theta_\xi-\frac{i}{2},
\end{gather}
\end{subequations}
one immediately obtains
\begin{align}\label{GeneralResult}
(\Delta\mathscr{M}_\pm(t))^2 &= \frac{1}{2}\left[\tilde{A}(t)\sin\left(\frac{\theta_\xi}{2}\right)-\tilde{B}(t)\cos\left(\frac{\theta_\xi}{2}\right)\right]^2e^{2r}\nonumber\\
&+ \frac{1}{2}\left[\tilde{A}(t)\cos\left(\frac{\theta_\xi}{2}\right)+\tilde{B}(t)\sin\left(\frac{\theta_\xi}{2}\right)\right]^2e^{-2r}\nonumber\\
&+ \frac{\kappa}{2}[\tilde{F}^2(t)+\tilde{G}^2(t)],
\end{align}
where
\begin{subequations}
\begin{align}
&\tilde{A}(t)\sin\left(\frac{\theta_\xi}{2}\right)-\tilde{B}(t)\cos\left(\frac{\theta_\xi}{2}\right) = A(t)\sin\left(\frac{\theta_\xi}{2}-\varphi\right)\label{AP1}\nonumber\\
&-B(t)\left[\frac{|\lambda|}{\Delta}\sin\left(\frac{\theta_\xi}{2}+\varphi^\prime\right)\pm\frac{\chi_s}{\Delta}\cos\left(\frac{\theta_\xi}{2}-\varphi\right)\right],\\
&\tilde{A}(t)\cos\left(\frac{\theta_\xi}{2}\right)+\tilde{B}(t)\sin\left(\frac{\theta_\xi}{2}\right) = A(t)\cos\left(\frac{\theta_\xi}{2}-\varphi\right)\nonumber\\
&-B(t)\left[\frac{|\lambda|}{\Delta}\cos\left(\frac{\theta_\xi}{2}+\varphi^\prime\right)\mp\frac{\chi_s}{\Delta}\sin\left(\frac{\theta_\xi}{2}-\varphi\right)\right].\label{AP2}
\end{align}
\end{subequations}
and
\begin{align}
\tilde{F}^2&(t)+\tilde{G}^2(t) = F^2(t)+2F(t)G(t)\frac{|\lambda|}{\Delta}\cos(\varphi+\varphi^\prime)\nonumber\\
&+G^2(t)\left(\frac{|\lambda|^2+\chi_s^2}{\Delta^2}\pm \frac{2|\lambda|\chi_s}{\Delta^2}\sin(\varphi+\varphi^\prime)\right).
\end{align}
Note that since $\varphi+\varphi^\prime = (\theta_\xi/2+\varphi^\prime)-(\theta_\xi/2-\varphi)$, one can see from Eq.\:\eqref{GeneralResult} that the noise of the output signals is dependent only on two phase differences, namely $\Delta\theta_1\equiv \theta_\xi/2-\varphi$ and $\Delta\theta_2\equiv \theta_\xi/2+\varphi^\prime$.

\subsection{The SNR for the cases with $\lambda_y=0$ and $\theta_\xi=\pi$}
Particularly, for the case with $\lambda_y=0$, $\theta_\xi=\pi$ and $\varphi=\pi/2$, the standard derivations of the time-integrated output quadratures can be simplified as
\begin{equation*}
(\Delta \mathscr{M}^\pm(t))^2=\frac{e^{-2r}}{2}A^{\prime 2}(t)+\frac{e^{2r}}{2}B^{\prime 2}(t)+\frac{\kappa}{2}(F^{\prime 2}(t)+G^{\prime 2}(t)),
\end{equation*}
where $A'(t)$, $B'(t)$, $F'(t)$, $G'(t)$ are determined by Eqs.\:\eqref{Aprime1} -- \eqref{Aprime2}. In such a scenario, the SNR becomes
\begin{equation*}
\mbox{SNR}(t)=\frac{2|\alpha||B(t)|}{\sqrt{e^{-2r}A_\lambda^2(t)+e^{2r}B^2(t)+\kappa(F^2_\lambda(t)+ G^2(t))}},
\end{equation*}
where
\begin{subequations}
\begin{align}
A_\lambda(t) &\equiv \frac{\Delta}{\chi_s}A^\prime(t) = \frac{\Delta}{\chi_s} A(t) +\frac{\lambda_x}{\chi_s} B(t),\\
F_\lambda(t) &\equiv \frac{\Delta}{\chi_s}F^\prime(t) = \frac{\Delta}{\chi_s} F(t) -\frac{\lambda_x}{\chi_s} G(t).
\end{align}
\end{subequations}
Similarly, for the case with $\lambda_y=0$, $\theta_\xi=\pi$ and $\varphi=0$, one will obtain
\begin{equation*}
(\Delta \mathscr{M}^\pm(t))^2=\frac{e^{2r}}{2}A^{\prime 2}(t)+\frac{e^{-2r}}{2}B^{\prime 2}(t)+\frac{\kappa}{2}(F^{\prime 2}(t)+G^{\prime 2}(t)),
\end{equation*}
where $A^\prime(t)$, $B^\prime(t)$, $F^\prime(t)$, $G^\prime(t)$ are determined by Eqs.\:\eqref{Bprime1} -- \eqref{Bprime2}. In such a scenario, the SNR becomes
\begin{equation*}
\mbox{SNR}(t)=\frac{2|\alpha||B(t)|}{\sqrt{e^{2r}A_\lambda^2(t)+e^{-2r}B^2(t)+\kappa(F^2_\lambda(t)+ G^2(t))}},
\end{equation*}
\begin{subequations}
where
\begin{align}
A_\lambda(t) &\equiv \frac{\Delta}{\chi_s}A^\prime(t) = \frac{\Delta}{\chi_s} A(t) -\frac{\lambda_x}{\chi_s} B(t),\\
F_\lambda(t) &\equiv \frac{\Delta}{\chi_s}F^\prime(t) = \frac{\Delta}{\chi_s} F(t) +\frac{\lambda_x}{\chi_s} G(t).
\end{align}
\end{subequations}

\subsection{The Phase matching conditions}
From Eq.\:\eqref{GeneralResult} and Eqs.\:\eqref{AP1} -- \eqref{AP2}, it is evident that in the case of a relatively short readout time, where the condition $A(t)\gg B(t)$ holds, it is advantageous to choose the phase matching conditions as
\begin{equation}
\frac{\theta_\xi}{2}-\varphi = m\pi,\frac{\theta_\xi}{2}+\varphi^\prime = n\pi,
\end{equation}
where $\varphi^\prime\equiv\varphi-\theta_\lambda$, and $m$ and $n$ are arbitrary integers. As such, the phase of the squeezed state and the phase of the nonlinear strength are determined by the local oscillator phase as
 \begin{equation}
 \theta_\xi=2\varphi + 2m\pi,\theta_\lambda = 2\varphi+(m-n)\pi,
 \end{equation}

For a slightly longer readout time, it is advantageous to select the phase matching condition as
\begin{equation}
\frac{\theta_\xi}{2}-\varphi = \frac{1}{2}(2m+1)\pi,
\frac{\theta_\xi}{2}+\varphi^\prime = \frac{1}{2}(2n+1)\pi.
\end{equation}
This implies that the phase of the squeezed state and the phase of the nonlinear strength are determined by the local oscillator phase as
\begin{equation}
\theta_\xi=2\varphi+(2m+1)\pi,\theta_\lambda = 2\varphi + (m-n)\pi.
\end{equation}
In this context, the term that is proportional to $e^{2r}$ in the noise term $(\Delta \mathscr{M}^\pm(t))^2$ takes the form
\begin{equation}
\frac{1}{2}\left(A(t)(-1)^{m-n} - B(t)\frac{|\lambda|}{\Delta}\right)^2.
\end{equation}
This term vanishes only when the readout time fulfills the condition $\Delta A(t)=|\lambda|B(t)$, and when $m-n$ is an even integer. 

Finally, for a sufficiently long readout time, it remains advantageous to choose
\begin{equation}
\frac{\theta_\xi}{2}-\varphi = \frac{1}{2}(2m+1)\pi.
\end{equation}
But now, using the following limits for long times
\begin{equation}
A(t\rightarrow\infty) = -\cosh 2\delta\: t,
B(t\rightarrow\infty) = \sinh 2\delta\: t,
\end{equation}
where
\begin{equation}
\frac{\kappa}{2}\equiv\sqrt{\frac{\kappa^2}{4}-\Delta^2}\cosh\delta\:\:\mbox{and}\:\:\Delta\equiv \sqrt{\frac{\kappa^2}{4}-\Delta^2}\sinh\delta,
\end{equation}
the phase of the nonlinear strength is determined by
\begin{equation}
(-1)^m\cosh 2\delta +\sinh 2\delta \frac{|\lambda|}{\Delta}\sin\left(\frac{\theta_\xi}{2}+\varphi^\prime\right)=0,
\end{equation}
or equivalently
\begin{equation}\label{2ndphase}
\sin\left(\frac{\theta_\xi}{2}+\varphi^\prime+m\pi\right) = -\frac{\frac{\kappa^2}{4}+|\lambda|^2-\chi_s^2}{\kappa |\lambda|}.
\end{equation}
Eq.\:\eqref{2ndphase} has real solutions if and only if
\begin{equation}
-\chi_s+\frac{\kappa}{2}\leq |\lambda|\leq \chi_s+\frac{\kappa}{2}\:\:\mbox{and}\:\:|\lambda|\geq \chi_s-\frac{\kappa}{2}.
\end{equation}
When the above condition is fulfilled, one obtains the phase matching conditions in the long time limit
\begin{equation}
\frac{\theta_\xi}{2}+\varphi^\prime+m\pi = -\frac{\pi}{2}\pm\left[\frac{\pi}{2}-\arcsin\left(\frac{\frac{\kappa^2}{4}+|\lambda|^2-\chi_s^2}{\kappa |\lambda|}\right)\right].
\end{equation} 
For example, when $\kappa=2\chi_s$, one will obtain
\begin{equation}
\frac{\theta_\xi}{2}+\varphi^\prime+m\pi = -\frac{\pi}{2}\pm\left[\frac{\pi}{2}-\arcsin\left(\frac{|\lambda|}{2\chi_s}\right)\right].
\end{equation}
Hence, in the long time limit, the phase of the squeezed state and the phase of the nonlinear strength are determined by the local oscillator phase as
\begin{subequations}
\begin{align}
\theta_\xi&=2\varphi+(2m+1)\pi,\\
\theta_\lambda &= 2\varphi+\pi \pm \left[\frac{\pi}{2}-\arcsin\left(\frac{\frac{\kappa^2}{4}+|\lambda|^2-\chi_s^2}{\kappa |\lambda|}\right)\right].
\end{align}
\end{subequations}

\subsection{Measurement time $t_c$ at which $A_\lambda(t_c)=0$}
Notably, in the latter scenario where $\lambda_y=0$, $\theta_\xi=\pi$ and $\varphi=0$, it is possible to significantly improve the SNR at the measurement time $t_c$ when $A_\lambda(t_c)=0$. To illustrate this, one may rewrite $A(t)$ and $B(t)$ as
\begin{align*}
A(t)&=\frac{\cosh 2\delta}{\Delta}(\sinh 2\delta- \tau)-\frac{\sinh 2\delta}{\Delta}e^{-\frac{\kappa}{2}t}\cosh(\tau+2\delta),\\
B(t)&= \frac{\sinh 2\delta}{\Delta}e^{-\frac{\kappa}{2}t}\sinh(\tau+2\delta)-\frac{\sinh 2\delta}{\Delta}(\sinh 2\delta- \tau),
\end{align*}
where $\tau\equiv \Delta t$. It immediately yields
\begin{align}
&A_\lambda(t) = \frac{\Delta \cosh 2\delta+\lambda_x\sinh 2\delta}{\Delta\chi_s}(\sinh 2\delta- \tau)\nonumber\\
&-\frac{\sinh 2\delta}{\Delta\chi_s}e^{-\frac{\kappa}{2}t}(\Delta\cosh(\tau+2\delta)+\lambda_x\sinh(\tau+2\delta)).
\end{align}
For the case when $\lambda_x>0$, one can introduce an auxiliary angle $\Xi$ via $\lambda_x\equiv \chi_s\cosh\Xi$ and $\Delta\equiv \chi_s\sinh\Xi$, so that the expression of $A_\lambda(t)$ can be simplified as
\begin{equation*}
A_\lambda(t) = \frac{\sinh(2\delta+\Xi)}{\Delta}(\sinh 2\delta- \tau) -\frac{\sinh 2\delta}{\Delta}e^{-\frac{\kappa}{2}t}\sinh(\tau+2\delta+\Xi)
\end{equation*}
Clearly, the auxiliary angle $\Xi$ is real only when $\Delta^2>0$, or equivalently $|\lambda|^2>\chi_s^2$. Interestingly, the conditions under which both auxiliary angles $\delta$ and $\Xi$ are real establish a bound of the nonlinear strength, i.e., $\chi_s^2<|\lambda|^2<\frac{\kappa^2}{4}+\chi_s^2$. In such a scenario, the equation for $A_\lambda(t)=0$ becomes
\begin{equation}\label{Lambert}
\left(1- \frac{\tau}{\sinh 2\delta}\right)e^{\coth\delta \tau} -  \frac{\sinh(\tau+2\delta+\Xi)}{\sinh(2\delta+\Xi)}=0.
\end{equation}
Evidently, $\tau_c=0$ is a trivial zero of Eq.\:\eqref{Lambert}. To find another non-trivial zero, one may approximate Eq.\:\eqref{Lambert} as
\begin{equation}\label{Lambert2}
a-b\tau = e^{c\tau},
\end{equation}
where
\begin{equation}
a = \frac{2\sinh(2\delta+\Xi)}{e^{2\delta+\Xi}},b=\frac{a}{\sinh 2\delta},c=1-\coth\delta.
\end{equation}
While obtaining Eq.\:\eqref{Lambert2}, we employed an approximation $e^{-(\tau+2\delta+\Xi)}\ll 1$, a condition which can be verified in later computations. Eq.\:\eqref{Lambert2} can be further reformulated into the standard form of the Lambert-$W$ equation as
\begin{equation}\label{Lambert3}
We^W = z,
\end{equation}
where
\begin{equation}
W\equiv\frac{ca}{b}-c\tau,z\equiv \frac{c}{b}e^{\frac{ca}{b}}.
\end{equation}
Eq.\:\eqref{Lambert3} is exactly solved by $W = W_\nu(z)$, or equivalently
\begin{equation}
\tau_c = \frac{a}{b}-\frac{1}{c}W_\nu\left(\frac{c}{b}e^{\frac{ca}{b}}\right).
\end{equation}
Here, $W_\nu(z)$ represents the Lambert-$W$ function, with $\nu=0$ or $-1$ denoting its two branches. When $-1/e\leq z<0$, there are two solutions $W=W_0(z)$ and $W=W_{-1}(z)$, namely the principal and negative branches, respectively. When $z>0$, there exists only the  principal branch $W_0(z)$, and the solution is given by $W=W_0(z)$. 

For example, when $\kappa=2\chi_s$ and $\lambda_x=1.2\chi_s$, one will obtain $\delta =0.7987$, $\Xi=0.6224$, $a=0.9882$, $b=0.4171$, and $c=-0.5076$, which yields $z=-0.3656$. As $-1/e=-0.3679\leq z<0$, there are two solutions $\tau_c=0.6103$ and $\tau_c=0.1710$. But the latter one is a false solution caused by neglecting the term $e^{-(\tau+2\delta+\Xi)}$. For the case when $\chi_s=2\pi\times 0.15\:\mbox{MHz}$, one will obtain $\Delta=0.6252\:\mbox{MHz}$, which yields $t_c=\tau_c/\Delta=0.9762\:\mu\mbox{s}$. Notably, when employing $\tau_c=0.6103$, $\delta =0.7987$, and $\Xi=0.6224$, one will obtain $e^{-(\tau+2\delta+\Xi)}\approx 0.0590\ll 1$, which confirms our initial approximation. Finally, solving Eq.\:\eqref{Lambert} numerically yields $\tau_c\approx 0.6566$. Hence, the error introduced by neglecting the term $e^{-(\tau+2\delta+\Xi)}$ is approximately $7\%$, which is tolerable. 

\section{Relation between fidelity and the signal-to-noise ratio}\label{B}
\subsection{Measurement fidelity for a qubit with an infinite life time}
To begin with, it is worthwhile to consider a continuous non-demolition measurement of a perfect qubit with an infinite life time, i.e., the qubit can not decay from the initial state. When the qubit is in the excited ($\sigma=+1$) or ground state ($\sigma=-1$), the probability distributions for the time-integrated output signal are the familiar Gaussian distributions
\begin{equation}\label{Gaussian}
P_{\pm}(x)=\frac{1}{\sqrt{2\pi}\sigma_{\pm}}\exp\left(\frac{-(x-\mu_{\pm})^2}{2\sigma_{\pm}^2}\right),
\end{equation}
where $x_\pm(t)\equiv \int_0^t P^\pm_{\mathrm{out}}(s)ds$ are the output signals integrated over time $t$, $\mu_{\pm}(t)\equiv \langle x_{\pm}(t)\rangle$ and $\sigma_\pm(t)\equiv \Delta x_\pm(t)$ are the mean and the standard derivation of $x_\pm(t)$ respectively. Without loss of generality, one can always assume $\mu_+\geq \mu_-$. As the distributions Eq.\:\eqref{Gaussian} are not necessary symmetric about $x=0$, one needs to set a signal threshold $\nu\equiv \frac{1}{2}(\mu_++\mu_-)$, and infers the measurement outcomes with $x>\nu$ as excited states, and those with $x<\nu$ as ground states. By definition, the fidelity is the difference between unity and the total probability for a false measurement inference
\begin{align}\label{FidelityDef}
F\equiv &1- P_+(x\leq \nu)- P_-(x\geq \nu)\nonumber\\
=&1  -\int_{-\infty}^\nu P_+(x)dx -\int_{\nu}^{\infty} P_-(x)dx.
\end{align}
Using Eq.\:\eqref{Gaussian}, a direct computation yields
\begin{align}\label{Pplus}
 P_+(x\leq \nu) &= \int_{-\infty}^\nu P_+(x)dx \nonumber\\
 &=\frac{1}{\sqrt{2\pi}\sigma_{+}}\int_{-\infty}^\nu\exp\left(\frac{-(x-\mu_{+})^2}{2\sigma_{+}^2}\right)dx \nonumber\\
 &=\frac{1}{\sqrt{\pi}}\int_{-\infty}^{\frac{\nu-\mu_+}{\sqrt{2}\sigma_+}}e^{-s^2}ds,
\end{align}
where $s\equiv (x-\mu_+)/(\sqrt{2}\sigma_+)$. A similar computation yields
\begin{equation}\label{Pminus}
P_-(x\geq \nu) = \frac{1}{\sqrt{\pi}}\int_{\frac{\nu-\mu_-}{\sqrt{2}\sigma_-}}^{\infty}e^{-s^2}ds.
\end{equation}
Substitution of Eqs.\:\eqref{Pplus} and \eqref{Pminus} into Eq.\:\eqref{FidelityDef} will yield
\begin{align}
F &= \frac{1}{\sqrt{\pi}}\int_{\frac{\nu-\mu_+}{\sqrt{2}\sigma_+}}^{\frac{\nu-\mu_-}{\sqrt{2}\sigma_-}}e^{-s^2}ds\nonumber\\
&= \frac{1}{2}\left[\erf\left(\frac{\mu_+-\mu_-}{2\sqrt{2}\sigma_-}\right)-\erf\left(\frac{\mu_--\mu_+}{2\sqrt{2}\sigma_+}\right)\right].
\end{align}
In particular, when the parameters are chosen so that the standard derivations of $x_+(t)$ and $x_-(t)$ are the same, i.e., $\sigma_+=\sigma_- = \sigma$, then one immediately obtains the fidelity
\begin{equation}
F = \erf\left(\frac{\mbox{SNR}}{\sqrt{2}}\right),
\end{equation}
where $\mbox{SNR}\equiv (\mu_+-\mu_-)/(2\sigma)$ is the signal-to-noise ratio for the scenario for $\sigma_+=\sigma_-$.

\subsection{Measurement fidelity for a qubit with a finite life time}
We now consider the measurement fidelity for a qubit with a finite life time. Unlike a perfect qubit, an imperfect qubit, when initially prepared in the excited state, can decay into the ground state at a later time $t_d$, which yields a jump of the output signal at the same time. By contrast, a qubit which is initially prepared in the ground state will not experience any excitations at later times. The decay time itself has to obey the exponential distribution: $P(t_d)=(1/T_1)\exp(-t_d/T_1)$, where $T_1\equiv\langle t_d\rangle$ is the single-qubit relaxation time. Hence, the time-integrated output signal for the initially excited state, a function of both the measurement time and the exponentially distributed relaxation time $t_d$ is explicitly given by
\begin{align}\label{SignalFiniteLife}
\tilde{x}_+(t) &= \int_0^{t_d} p_{\mathrm{out}}^+(s)ds + \int_{t_d}^t p_{\mathrm{out}}^-(s)ds\nonumber\\
&=\delta x(t_d)+x_-(t),(t\geq t_d),
\end{align}
where $\delta x(t_d)\equiv x_+(t_d)-x_-(t_d)$ is the output signal difference of a perfect qubit which is initially in the excited or ground states at the relaxation time $t_d$. From Eq.\:\eqref{SignalFiniteLife}, one obtains the mean of the output signals: $\tilde{\mu}_+(t)\equiv \langle \tilde{x}_+(t) \rangle=\delta\mu(t_d)+\mu_-(t)$, where $\delta\mu(t_d)\equiv \mu_+(t_d)-\mu_-(t_d)$. Hence, the time-integrated output signal for an imperfect qubit at any time before or after $t_d$, is given by
\begin{equation}
\tilde{x}_+(t) = x_+(t)\theta(t_d-t)+(\delta x(t_d)+x_-(t))\theta(t-t_d),
\end{equation}

Hence, when the qubit is initially prepared in the excited state, the conditional probability distribution for the time-integrated output signal given a decay time $t_d$ has the form
\begin{align}
P_+&(x|t_d) = \frac{1}{\sqrt{2\pi}\sigma_+}\left\{\exp\left(-\frac{(x-\mu_+)^2}{2\sigma_+^2}\right)\theta(t_d-t)\right.\nonumber\\
&\left.+ \exp\left[-\frac{(x-\delta_\mu(t_d)-\mu_-)^2}{2\sigma_+^2}\right]\theta(t-t_d)\right\}.
\end{align}
Averaging over all possible decay times, one obtains the unconditional distribution for the time-integrated output signal
\begin{align}\label{UnCon}
&P_+(x) = \frac{1}{\sqrt{2\pi}\sigma_+}\left\{\int_t^{\infty}\exp\left(-\frac{(x-\mu_+)^2}{2\sigma_+^2}\right)\frac{e^{-\frac{t_d}{T_1}}}{T_1}dt_d\right.\nonumber\\
&\left.+\int_0^t\exp\left[-\frac{(x-\delta_\mu(t_d)-\mu_-)^2}{2\sigma_+^2}\right]\frac{e^{-\frac{t_d}{T_1}}}{T_1}dt_d\right\}.
\end{align}
The first integral of Eq.\:\eqref{UnCon} is easy to evaluate, which is simply the Gaussian distribution multiplied by an exponential decay factor $\exp(-t/T_1)$. In terms of the dimensionless decay time $\tau_d\equiv t_d/T_1$, the unconditional distribution $P_+(x)$ becomes
\begin{align}\label{GeneralPplus}
&P_+(x) = \frac{1}{\sqrt{2\pi}\sigma_+}\left\{\exp\left(-\frac{(x-\mu_+)^2}{2\sigma_+^2}-\frac{t}{T_1}\right)\right.\nonumber\\
&\left.+\int_0^{t/T_1}\exp\left[-\frac{(x-\mu_--\delta_\mu(T_1\tau_d))^2}{2\sigma_+^2}-\tau_d\right]d\tau_d\right\}.
\end{align}
In particular, when the separation of the average values of signals is a linear function of time, i.e., $\mu_+(t)=-\mu_-(t)= ct$, the remaining integral is a Gaussian integral, which can be solved to obtain the final expression of the unconditional distribution
\begin{align}
&P_+(x) = \frac{1}{\sqrt{2\pi}\sigma_+}\left\{\exp\left(-\frac{(x-\mu_+)^2}{2\sigma_+^2}-\frac{t}{T_1}\right)\right.\nonumber\\
&+\frac{1}{2\xi}\exp\left(\frac{\sigma_+^2}{2\xi^2}-\frac{x+ct}{\xi}\right)\nonumber\\
&\cdot\left[\erf\left(\frac{\sigma_+}{\sqrt{2}\xi}-\frac{x-ct}{\sqrt{2}\sigma_+}\right)-\erf\left(\frac{\sigma_+}{\sqrt{2}\xi}-\frac{x+ct}{\sqrt{2}\sigma_+}\right)\right],
\end{align}
where $\xi\equiv 2cT_1$ is the separation of the average values of the signals at $T_1$.

One can see from Eq.\:\eqref{GeneralPplus} that when $T_1\rightarrow\infty$, the unconditional distribution $P_+(x)$ becomes the familiar Gaussian distribution for a perfect qubit. When the measurement time is much less than the single-qubit relaxation time, i.e., $t\ll T_1$, one can approximate the second integral in Eq.\:\eqref{GeneralPplus} by using the trapezoidal rule, so that
\begin{align}\label{trapezoidal}
&\int_0^{t/T_1}\exp\left[-\frac{(x-\mu_--\delta_\mu(T_1\tau_d))^2}{2\sigma_+^2}-\tau_d\right]d\tau_d\approx \frac{t}{2T_1}\nonumber\\
&\cdot\left\{\exp\left(-\frac{(x-\mu_-)^2}{2\sigma_+^2}\right)+e^{-t/T_1}\exp\left(-\frac{(x-\mu_+)^2}{2\sigma_+^2}\right)\right\}
\end{align}
where we have used the fact that $\mu_-+\delta\mu =\mu_+$. Substitution of Eq.\:\eqref{trapezoidal} into Eq.\:\eqref{FidelityDef}, one immediately obtains the fidelity
\begin{align}
&F \equiv 1 - \int_\nu^\infty P_-(x)dx-\int_{-\infty}^\nu P_+(x)dx\nonumber\\
=&\frac{1}{2}\left[1+\erf\left(\frac{\nu-\mu_-}{\sqrt{2}\sigma_-}\right)\right]-\frac{e^{-t/T_1}}{2}\left[1+\erf\left(\frac{\nu-\mu_+}{\sqrt{2}\sigma_+}\right)\right]\nonumber\\
-&\frac{t}{4T_1}\left\{1+\erf\left(\frac{\nu-\mu_-}{\sqrt{2}\sigma_+}\right)+e^{-t/T_1}\left[1+\erf\left(\frac{\nu-\mu_+}{\sqrt{2}\sigma_+}\right)\right]\right\},
\end{align}
where $\nu$ is the signal threshold determined by the condition $P_+(\nu)=P_-(\nu)$. For the symmetric case when the standard derivations $\sigma_+$ and $\sigma_-$ are the same, the signal threshold is still given by $\nu=(\mu_++\mu_-)/2$, provided that the measurement time $t$ is much less than the single-qubit relaxation $T_1$. In such a case, the fidelity can be expressed in terms of the signal-to-noise ratio as 
\begin{align}
F &= \frac{1}{2}\left(1-\frac{t}{2T_1}\right)\left[1+\erf\left(\frac{\mbox{SNR}}{\sqrt{2}}\right)\right]\nonumber\\
&-\frac{e^{-t/T_1}}{2}\left(1+\frac{t}{2T_1}\right)\left[1-\erf\left(\frac{\mbox{SNR}}{\sqrt{2}}\right)\right].
\end{align}
Finally, using the approximation $e^{-t/T_1}\approx 1-t/T_1$ and ignoring the second-order terms in $t/T_1$, one obtains the following expression for the fidelity
\begin{equation}
F \approx \exp\left(-\frac{t}{2T_1}\right)\erf\left(\frac{\mbox{SNR}}{\sqrt{2}}\right). 
\end{equation}

\section{Squeezing: State-of-the-art-Technologies}
In the literatures, the strength of squeezing is usually measured in terms of the squeezing factor $R$ in dB units, where the connection between the squeezing factor $R$ and the squeezing parameter $r$  used in our manuscript is given by
\begin{equation}
R = -10\log_{10}(e^{-2r}).
\end{equation}
For example, a squeezing factor of $11.35$ dB corresponds to a squeezing parameter $r\approx 1.31$, a squeezing factor of $15$ dB corresponds to a squeezing parameter $r\approx 1.73$, and a squeezing factor of $20$ dB corresponds to a squeezing parameter $r\approx 2.3$. Conversely, a squeezing parameter $r=2$ corresponds to a squeezing factor $R\approx 17.37$ dB, and a squeezing parameter $r=2.5$ corresponds to a squeezing factor $R\approx 21.71$ dB.

In current state-of-the-art experiments, the squeezing factor is limited to a few decibels or several dozens of decibels depending on different squeezing mechanisms. For instance, in the latest observing run of the advanced laser interferometer gravitational wave experiments, a squeezed state of light at 1064 nm with a squeezing factor 2.7 dB is employed \cite{tse2019quantum}. Using a positive intrinsic negative ($p-i-n$) photodiode, a direct observation of squeezed vacuum state of light with a squeezing factor of up to 15 dB was demonstrated at a wavelength of 1064 nm \cite{vahlbruch2016detection}. Besides, a squeezed state of microwave radiation with a squeezing factor of up to 8 dB has been reported using mechanical oscillator \cite{ockeloen2017noiseless}. Similarly, a broadband squeezed microwave radiation with a squeezing factor of up to 11.35 dB for a single-mode field, and a squeezing factor of up to 9.54 dB for a two-mode field has been demonstrated over a bandwidth of 1.75 GHz \cite{qiu2022broadband}. 

\end{appendix}

\end{document}